\documentclass[preprint,1p,times,twocolumn]{elsarticle}
\usepackage{natbib}
 \usepackage{graphicx}

\usepackage{amssymb}

\biboptions{comma,round}

\journal{Nuclear Physics A}

\begin{document}

\begin{frontmatter}


\title{Elliptic flow of charged pions, protons and strange particles
  emitted in Pb+Au collisions at top SPS energy\tnoteref{t1}}
\tnotetext[t1]{CERES Collaboration}
\author[a]{D.~Adamov{\'a}}
\author[b,k]{G.~Agakichiev}
\author[c,l]{A.~Andronic}
\author[d]{D.~Anto{\'n}czyk}
\author[d]{H.~Appelsh{\"a}user}
\author[b]{V.~Belaga}
\author[e,f,m]{J.~Biel\v{c}\'{\i}kov{\'a}}
\author[c,l]{P.~Braun-Munzinger}
\author[f]{O.~Busch}
\author[g]{A.~Cherlin}
\author[f]{S.~Damjanovi{\'c}}
\author[h]{T.~Dietel}
\author[f]{L.~Dietrich}
\author[i]{A.~Drees}
\author[f]{W.~Dubitzky}
\author[f,n]{S.~I.~Esumi}
\author[f,o]{K.~Filimonov}
\author[b]{K.~Fomenko}
\author[g]{Z.~Fraenkel\corref{cor1}}
\author[c]{C.~Garabatos}
\author[f]{P.~Gl{\"a}ssel}
\author[c]{G.~Hering}
\author[c]{J.~Holeczek}
\author[c]{M.~Kalisky}
\author[f]{G.~Krobath}
\author[a]{V.~Kushpil}
\author[c]{A.~Maas}
\author[c,l]{A.~Mar\'{\i}n}
\author[f,p]{J.~Milo\v{s}evi{\'c}}
\ead{Jovan.Milosevic@cern.ch}
\author[c,l]{D.~Mi{\'s}kowiec}
\author[b]{Y.~Panebrattsev}
\author[b]{O.~Petchenova}
\author[f,q]{V.~Petr{\'a}\v{c}ek}
\author[c]{S.~Radomski}
\author[c,r]{J.~Rak}
\author[g]{I.~Ravinovich}
\author[j]{P.~Rehak\corref{cor1}}
\author[c]{H.~Sako}
\author[f]{W.~Schmitz}
\author[d]{S.~Schuchmann}
\author[c]{S.~Sedykh}
\author[b]{S.~Shimansky}
\author[f]{J.~Stachel}
\author[a]{M.~\v{S}umbera}
\author[f]{H.~Tilsner}
\author[g]{I.~Tserruya}
\author[c]{G.~Tsiledakis}
\author[h]{J.\thinspace P.~Wessels}
\author[f]{T.~Wienold}
\author[e]{J.\thinspace P.~Wurm}
\ead{J.P.Wurm@mpi-hd.mpg.de}
\author[f,s]{S.~Yurevich}
\author[b]{V.~Yurevich}
\cortext[cor1]{\it deceased}
\address[a]{Nuclear Physics Institute, Academy of Sciences of the Chech
Republic, 25068 \v{R}e\v{z}, Czech Republic}
\address[b]{Joint Institute of Nuclear Research, Dubna, 141980 Moscow Region, Russia}
\address[c]{Institut f\"ur Kernphysik, GSI, 64291 Darmstadt, Germany}
\address[d]{Institut f\"ur Kernphysik, Johann Wolfgang Goethe-Universit{\"a}t 
Frankfurt, 60438 Frankfurt, Germany}
\address[e]{Max-Planck-Institut f{\"u}r Kernphysik, 69117 Heidelberg, Germany} 
\address[f]{Physikalisches Institut, Universit{\"a}t Heidelberg, 69120 Heidelberg, Germany} 
\address[g]{Department of Particle Physics, Weizmann Institute, Rehovot, 76100 Israel} 
\address[h]{Institut f\"ur Kernphysik, Universit{\"a}t M{\"u}nster, 48149 M\"unster, 
Germany} 
\address[i]{Department for Physics and Astronomy, SUNY Stony
Brook, NY 11974, USA} 
\address[j]{Instrumentation Division, Brookhaven National Laboratory,
Upton, NY 11973-5000, USA} 
\address[k]{Present affiliation:~ II.~Physikalisches Institut der Justus Liebig 
Universit\"at, Giessen, Germany}
\address[l]{Present affiliation:~ Research Division and Extreme Matter Institute (EMMI),
GSI Helmholtzzentrum f\"ur Schwerionenforschung, 64291 Darmstadt, Germany}
\address[m]{Present affiliation:~ Nuclear Physics Institute, Academy of Sciences of 
the Czech Republic, 25068 \v{R}e\v{z}, Czech Republic}
\address[n]{Present affiliation:~ Institute of Physics, University of Tsukuba, Tsukuba, 
Japan}
\address[o]{Present affiliation:~ Physics Department, University of California, 
Berkeley, CA 94720-7300, USA}
\address[p]{Present affiliation:~ Faculty of Physics and Vin$\check{c}$a Institute of 
Nuclear Sciences, University of Belgrade,\\ 11001 Belgrade, Serbia}
\address[q]{Present affiliation:~ Faculty of Nuclear Science and Engineering,
Czech Technical University, Prague, Czech Republic}
\address[r]{Present affiliation:~ Department of Physics, University of Jyv\"askyl\"a,
Jyv\"askyl\"a, Finland}
\address[s]{Present affiliation:~ Institut f\"ur Kernphysik, GSI, 64291 Darmstadt, 
Germany}
\begin{abstract}
Differential elliptic flow spectra $v_2(p_T)$ of $\pi^{-}$,
$K^{0}_{S}$, $p$, $\Lambda$ have been measured at $\sqrt{s_{NN}}$=
17.3~GeV around midrapidity by the CERN-CERES/NA45 experiment in
mid-central Pb+Au collisions (10\% of $\sigma_{geo}$). The $p_T$ range
extends from about 0.1~GeV/c (0.55 GeV/c for $\Lambda$) to more than
2~GeV/c.  Protons below 0.4~GeV/c are directly identified by
$dE/dx$. At higher $p_T$, proton elliptic flow is derived as a
constituent, besides $\pi^+$ and $K^+$, of the elliptic flow of
positive pion candidates. This retrieval requires additional inputs:
(i) of the particle composition, and (ii) of $v_2(p_T)$ of {\it
positive} pions. For (i), particle ratios obtained by NA49 are
adapted to CERES conditions; for (ii), the measured $v_2(p_T)$ of
{\it negative} pions is substituted, assuming $\pi^+$ and $\pi^-$
elliptic flow magnitudes to be sufficiently close.  The $v_2(p_T)$
spectra are compared to ideal-hydrodynamics calculations. In synopsis
of the series $\pi^{-}$ - $K^{0}_{S}$ - $p$ - $\Lambda$, flow
magnitudes are seen to fall with decreasing $p_T$ progressively even
below hydro calculations with early kinetic freeze-out
($T_f$=~160~MeV) leaving not much time for hadronic evolution. The
proton $v_2(p_T)$ data show a downward swing towards low $p_T$ with
excursions into negative $v_2$ values. The pion-flow isospin asymmetry
observed recently by STAR at RHIC, invalidating in principle our
working assumption, is found in its impact on proton flow bracketed
from above by the direct proton flow data, and not to alter any of our
conclusions. Results are discussed in perspective of recent viscous
hydrodynamics studies which focus on late hadronic stages.
\end{abstract}


\end{frontmatter}

\pagenumbering{arabic}
\setcounter{page}{1}
\setcounter{section}{0}
\setcounter{figure}{0}

\section{Introduction}
\label{introduction}

Among the prominent results from the Relativistic Heavy Ion Collider
(RHIC) are observations of strong elliptic flow~\cite{bra05,pho05,
star05,phenix05} in non-central collisions characterised by
azimuthally anisotropic particle yields in the plane transverse to the
beam direction~\cite{Ollit92,Bar94, PosVol,Apel98}. Elliptic flow is
quantified by $v_2$, the second harmonic coefficient of the azimuthal
particle distribution with respect to the reaction plane. The
observations directly assert the importance of strong interactions
among constituents of the expanding, hot and dense medium by which the
geometrical anisotropy of the almond shaped overlap zone evolves into
the momentum space anisotropy that is measured. This evolution is
described by relativistic hydrodynamics~\cite{HuoRuu06}. More
specifically, the large $v_2$ values agreed surprisingly well with
predictions of hydrodynamics without dissipation. This was interpreted
as the early-time response of a locally equilibrated system of a very
peculiar kind, the strongly interacting Quark Gluon Plasma (QGP),
behaving as a nearly perfect liquid with a very small ratio $\eta/s$
of shear viscosity to entropy density~\cite{GyuMcL05,HHKLN06}.

From $\sqrt{s_{NN}}$= 200~GeV at RHIC to $\sqrt{s_{NN}}$= 2.76~TeV a
bold step upward in nucleon-nucleon centre-of-mass energy was recently
taken with the operation of the Large Hadron Collider (LHC). First
measurements of elliptic flow in Pb+Pb collisions report $v_2(p_T)$
for charged hadrons of similar magnitude and shape than at comparable
centralities at RHIC~\cite{Aamodt11,Atlas1108,Tserruya11}. The average
$v_2$ is $\approx$20\% larger at the LHC but this increase is mainly
due to the harder $p_T$ spectrum at LHC energies.  This is in
agreement with hydrodynamic predictions extrapolated from RHIC data
without change in the (very low) viscosity to entropy
density~\cite{Luzum11,SHHS11}, as well as a hybrid calculation
treating the QGP by ideal hydrodynamics and the late stages by a
hadronic cascade model~\cite{HHN10}.

At the Super Proton Synchrotron (SPS) energy, $\sqrt{s_{NN}}$=
17.3~GeV, elliptic flow magnitudes $v_2$ are about 30\% lower than at
RHIC. The differential flow data $v_2(p_T)$ at
SPS~\cite{NA4903,Aga04,Aggar04}, though strikingly similar in shape to
the RHIC and LHC data, stay well below calculations of ideal
hydrodynamics~\cite{Huo_plb01}.  It should be noted that even at the
highest RHIC energy some significant deviations remain~\cite{Heinz05}.

The failure of hydrodynamics at low energy, large impact parameters or
large forward rapidities has been ascribed to insufficient number
densities at very early collision stages that hamper
thermalization~\cite{Heinz04}. Strong dissipative effects are bound to
set in after chemical freeze-out with growing mean free paths during
the late hadronic expansion~\cite{GyuMcL05,HHKLN06,Teaney03,Niemi11}.

We present differential elliptic flow data $v_2(p_T)$ of of strange
particles $\Lambda$ and $K^{0}_{S}$ for mid-central 158~$A$GeV Pb+Au
collisions collected by the CERES/NA45 experiment at the
SPS~\cite{Jovan06}; of negative pions complementing earlier, more
peripheral data~\cite{Aga04,Jana03}; and of protons directly measured
at low $p_T$ by $dE/dx$ identification and retrieved at higher $p_T$
from the measured elliptic flow data of positive pion candidates
containing besides pions and protons also kaons. The latter task
assumes charge-independent pion flow, at least to an accuracy allowing
to substitute the {\it negative}-pion $v_2(p_T)$ measured for the {\it
positive}-pion $v_2(p_T)$ required. The particle composition also
needed for the retrieval is fixed by particle ratios from
the NA49 Collaboration that are adapted to CERES conditions.

Elliptic flow data of identified particles as presented here are
sparse at the SPS, especially for massive particles, and reaching down
to low $p_T$. We foresee these data to contribute valuable information
on late stages of collective expansion which has been characterized by
rescattering in the `hadronic corona'~\cite{HirGyu_NPA06} as an
interplay between strong radial flow, elliptic flow, and thermal
motion set by the freeze-out temperature~\cite{Huo_plb01}. The proton
flow data may turn out worth our effort as a probe specifically
sensitive to hadronic viscosity~\cite{Shen11}.

The assumed pion-flow isospin symmetry has recently been found
invalidated by results of the `beam energy scan' of the STAR
Collaboration~\cite{bes11}, in minimum-bias collisions at low
$\sqrt{s_{NN}}$ and low ${p_T}$. The yet unknown centrality dependence
of the effect will be discussed in view of diverse physics scenarios.
As a worst case remedy, our {\it directly} measured proton-flow data
provide an upper limit for the uncertainty in proton $v_2$ inflicted
by the pion-flow asymmetry.

\thispagestyle{empty}
\cleardoublepage 

\section{Experiment}
\label{experiment}

\begin{figure}[h]
\centerline{\includegraphics[width=10cmp]{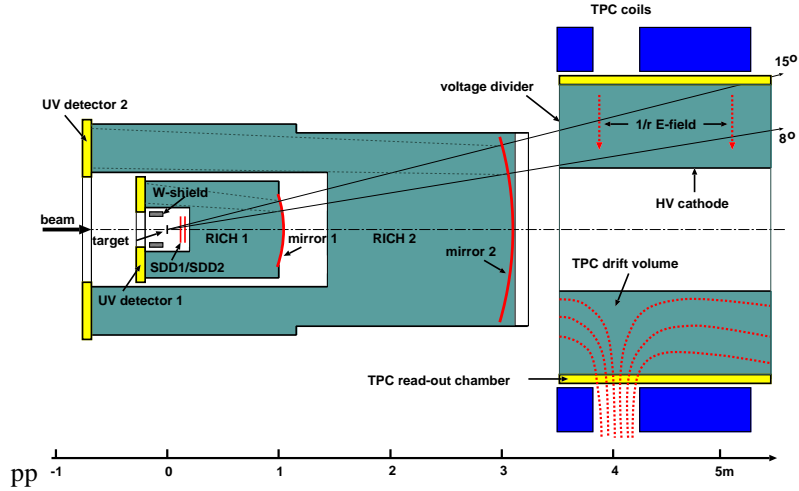}}
  \caption{The CERES/NA45 spectrometer during final data taking in 2000. 
  \label{fig:exp}}
  \end{figure}

The 158 $A$GeV Pb+Au data were collected with the upgraded
CERES/NA45 spectrometer during the heavy-ion run in 2000 at the CERN
SPS. The CERES spectrometer is axially symmetric around the beam
direction and covers full azimuth at polar angles 7.7$^\circ \leq
\vartheta \leq$ 14.7$^\circ$, corresponding to a pseudorapidity range
$2.05< \eta < 2.70$ close to midrapidity ($y_{mid}$=\,2.91); it is
thus very well suited for elliptic flow studies. A cross-section view
of the spectrometer is shown in Fig.~\ref{fig:exp}. A detailed
description of the CERES experiment is given in \cite{Marin04}.

The radial-drift Time Projection Chamber (TPC)~\cite{NIM08} is
operated inside a magnetic field with maximum radial component of
0.5~T, providing a precise determination of the momentum. Particle
identification is achieved by the differential energy loss d$E$/d$x$ along
the tracks in the TPC. A doublet of radial Silicon Drift Detectors
(SDD)~\cite{Holl96}, located at 10 and 13~cm downstream of a segmented
Au target, is used for vertex reconstruction and tracking outside the
field region. Charged particles emitted from the target are
reconstructed by matching track segments in the SDD and in the TPC
using a momentum-dependent matching window. The two Ring Imaging
Cherenkov Counters (RICH1, RICH2) for electron identification were
used in a previous CERES flow study of identified charged
pions~\cite{Aga04}, but not in the measurement reported on here.

\subsection{Momentum Resolution}
\label{momres}

The momentum is measured by determining the deflection of the tracks
within the TPC. The momentum resolution is therefore depending
on the spatial track resolution, but is degraded by multiple
scattering in the detector material. 

The results of an extensive Monte-Carlo study of the detector response 
were shown~\cite{yurevich06} to be well approximated by the 
simple expression
\begin{equation}
\Delta p/p= \sqrt{(2.0\%)^2+ (1.0\% \cdot p[{\rm GeV/c}])^2}. 
\end{equation}

\subsection{Trigger Samples}
\label{ident}

\begin{figure}[b!]
  \centerline{\includegraphics[width=9.5cm]
{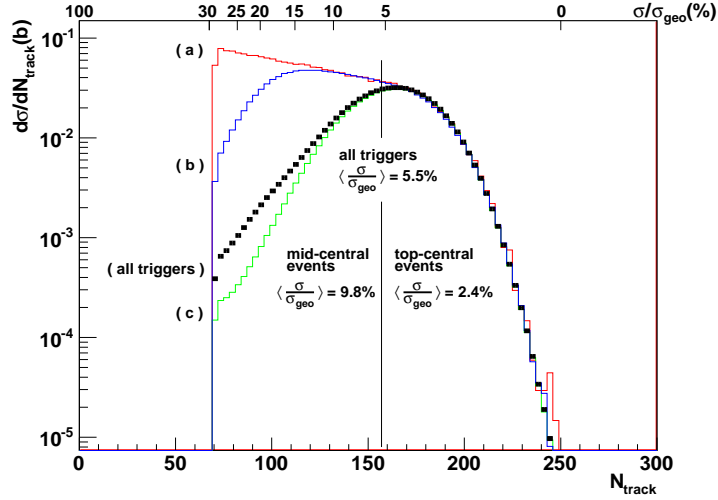}}
  \caption{TPC track density for various trigger selections, (a)
 minimum-bias, (b) peripheral, (c) central. The combination of all
 triggers (`all triggers'), dominated by (c), has a weighted mean
 $\langle\sigma/\sigma_{geo}\rangle$ of 5.5~\%.  It is split into
 `mid-central' triggers (left, $N_{track}<$ 159) of 9.8~\% by which
 most of the flow data were collected; and `top-central' triggers
 (right, $N_{track}\geq$ 159 ) of 2.4~\%, respectively.
 The horizontal scale $\langle\sigma/\sigma_{geo}\rangle$ on top applies
 to (a) only. See text.
\label{fig:mult}}
  \end{figure}

A sample of 30$\cdot 10^{6}$ events of 158~$A$GeV Pb+Au collisions was
collected by a mixed-trigger selection with average centrality
$\sigma/\sigma_{geo}$= 5.5\%; this choice was made to enhance $e^+e^-$
production, CERES' main objective. The track-multiplicity distribution
for `all triggers', shown in Fig.~\ref{fig:mult} by squares, has an
average track number $<\!\!N_{track}\!\!>_{|_{all}}$= 157.9. It
strongly deviates at low multiplicities from the minimum-bias
distribution labeled $(a)$ in Fig.~\ref{fig:mult}: for the same average
multiplicity, the minimum-bias distribution would have to be cut at
$N_{track}$= 129.1 or at the top 11.4\% of $\sigma_{geo}$.

The limited statistics of our strange-particle spectra allowed for
only two centrality classes. The `top-central' part matches the
minimum-bias distribution almost to its cut off at $N_{track}$=~159
and comprises with $<\!\!N_{track}\!\!>_{|_{top}}$= 176.9 the top
central 2.4\% of $\sigma_{geo}$.  The remainder are `mid-central'
triggers, $<\!\!N_{track}\!\!>_{|_{mid}}$= 136.2; these were used
almost exclusively to collect the elliptic flow data. A precise
definition of these triggers can only be provided by the distribution
itself, together with the minimum-bias curve. For comparison to other
experiments, we quote the slice cut from the minimum-bias distribution
that has identical average multiplicity: it extends from 5.3\% to
14.5\% of $\sigma_{geo}$ with a weighted average
$\langle\sigma/\sigma_{geo}\rangle= 9.8\%$.

\subsection{Pion and Proton Identification}
\label{pionident}

Tracks in the TPC were reconstructed in the pseudorapidity range 
2.05$\leq \eta \leq$ 2.70. They had to pass quality cuts that 
required

{\itemize  
\item transverse momentum $p_{T}$ above 50~MeV/c;
\item a minimum of 8 to 12 hits per track, depending on polar angle 
$\vartheta$;
\item TPC and SDD tracks of charged pions and primary protons to match within 
a $\pm$3$\sigma$ $p_T$-dependent window.  
\enditemize}

 \begin{figure}[t!]
  \centerline{\includegraphics[width=13.0cm,height=9.0cm]
  {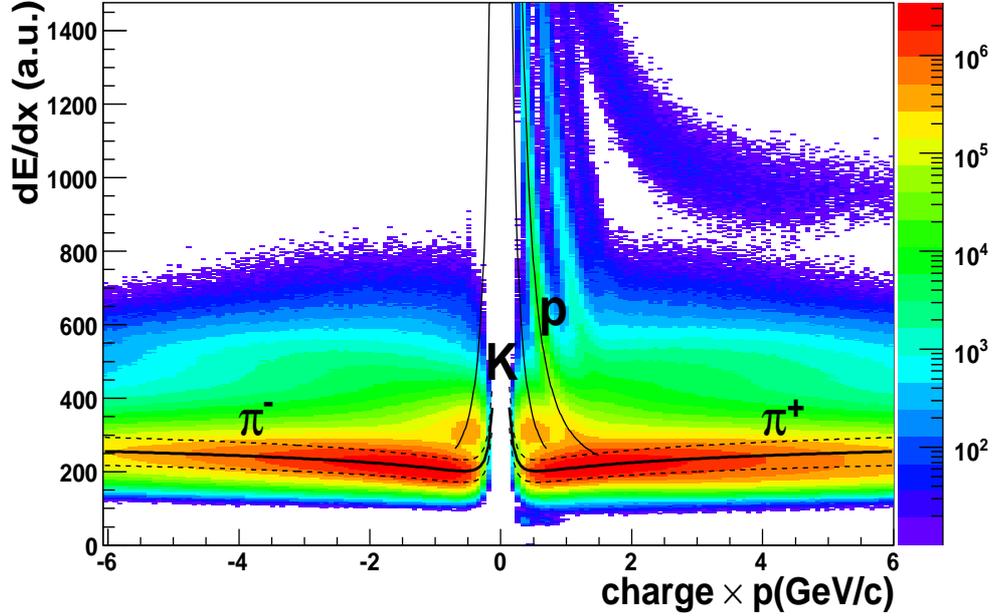}}
  \caption{Contour plots of the specific energy loss of
   charged particles {\it vs} momentum times charge sign. 
   Full lines show Bethe-Bloch energy loss for $\pi$, $K$, $p$, dashed 
   lines the $\pm 1.5\sigma$ cut selecting pions. Relative d$E$/d$x$ 
   resolution, depending on number of hits per track, is typically 10\%.
   \label{fig:dEdx} }
\end{figure}
Pion candidates are identified by their specific-energy loss sampled
along the tracks in the TPC. On average, there are more than 10 hits
per track. 

In the two-dimensional scatter plot of Fig.~\ref{fig:dEdx}, the
measured specific energy loss ~d$E$/d$x$~ is shown as function of
particle momentum $p$~ for both negative and positive charges.  Shown
are the cuts of Eq.~\ref{eq:contour} selecting pion candidates in a
$\pm 1.5\sigma$, i.e. $\pm$15\%, window (dashed lines) around the
nominal energy loss for charged pions according to the Bethe-Bloch
(BB) formula (full lines). The d$E$/d$x$ cut is defined as
\begin{equation}
\label{eq:contour}
0.85 \frac{{\rm d}E}{{\rm d}x}(p,\pi^{\pm})\mid_{\rm BB} \le \frac{{\rm d}E}
{{\rm d}x}(p)\mid_{\rm measured} \le 1.15 \frac{{\rm d}E}{{\rm d}x}
(p,\pi^{\pm})\mid_{\rm BB}.
\end {equation} 

\begin{figure}[t!]
  \centerline{\includegraphics[width=12.2cm]
  {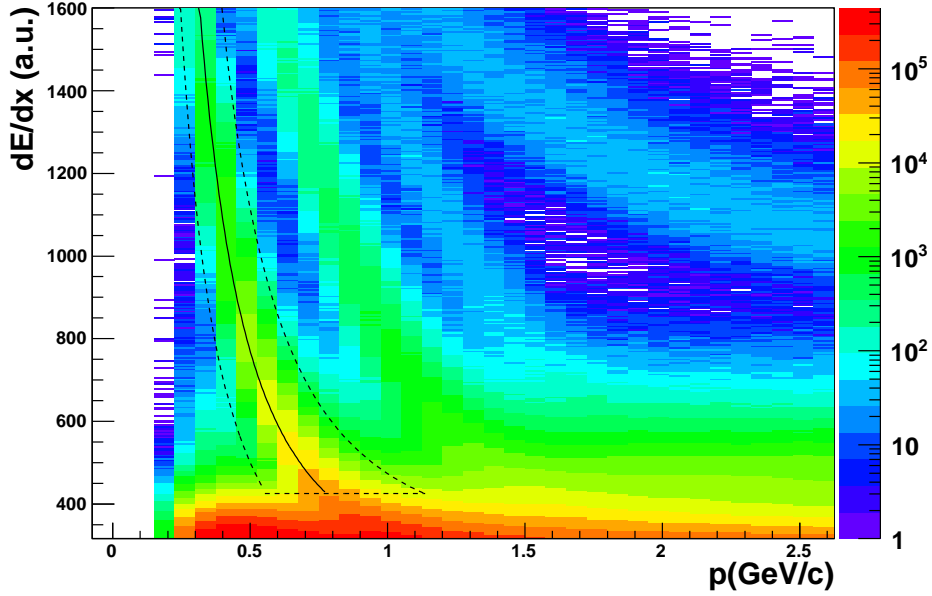}}
  \caption{The low momentum part of Fig.~\ref{fig:dEdx} for positive charges
  with lines indicating the cut to select protons.
\label{fig:ident_dEdx}}
  \end{figure}

Also shown are the Bethe-Bloch lines for kaons and protons, and it is
obvious that over extended ranges in momentum pions will be mixed with
kaons and protons, in case of positive charge. Antiprotons are only
about 6\% of protons at mid-rapidity.  At very low momenta protons are
well identified by d$E$/d$x$ as can be observed from
Fig.~\ref{fig:ident_dEdx}.

\subsection{Reconstruction of $\Lambda$ and $K_{S}^{0}$}
\label{reco}

The $\Lambda$ particles are reconstructed via the decay channel
$\Lambda \rightarrow p+\pi^{-}$ with branching ratio $BR=63.9\%$ and
mean decay length $c\tau$=~7.89~cm~\cite{PDG04}. Particle identification
is performed using the d$E$/d$x$ samples from the TPC by applying a $\pm
1.5\sigma$ and $\pm 1.0\sigma$ window around the momentum dependent
 \begin{figure}[t!]
  \centerline{\includegraphics[width=12.5cm,height=8.5cm]
 {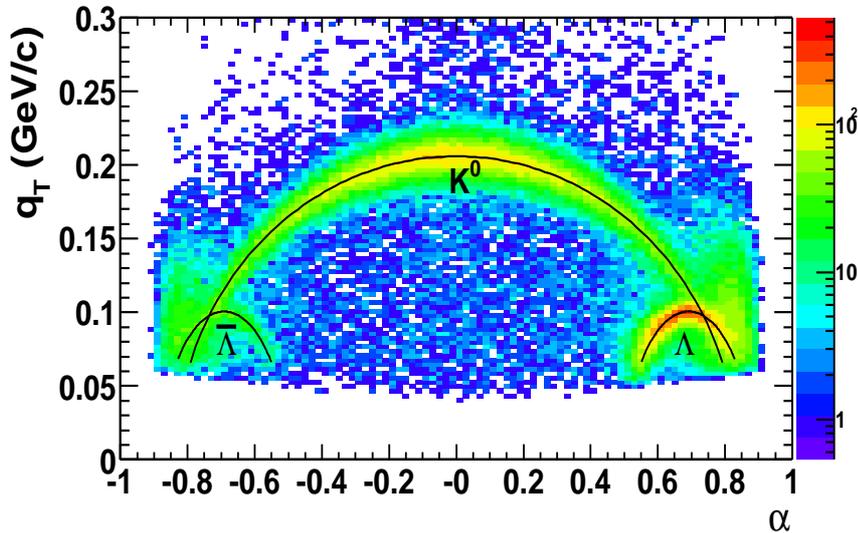}}
  \caption{Armenteros-Podolanski plot shows {\it loci} of $\Lambda$,
   $\bar{\Lambda}$ and $K_{S}^{0}$ reconstructed from experimental
   data; taken from \cite{Wilrid}. See text.
 \label{fig:arm_Lambda_Kaon}}
 \end{figure}
Bethe-Bloch values for pions and protons, respectively. The long decay
length allows to accept only $\Lambda$ decays which have no match to a
SDD track within $\pm 3\sigma$ window. On the pair level, a $p_{T}$
dependent opening angle cut $\vartheta_{p\pi^{\pm}}$ is applied. In
addition, an Armenteros-Podolanski cut~\cite{PodAr54} with
$q_T\leq$0.125~GeV/c and 0.0$\leq\alpha\leq$~0.65 is applied in order
to suppress background, admittedly with a considerable loss of signal.
The $\alpha$ variable is a measure of the longitudinal momentum
asymmetry, $\alpha=(q^{+}_{L}-q^{-}_{L})/(q^{+}_{L}+q^{-}_{L})$, where
$q^{+}_{L}$ and $q^{-}_{L}$ denote longitudinal momentum components
of $\vec{p}^{+}$ and $\vec{p}^{-}$ calculated with respect to
$\vec{p}_{\Lambda}=\vec{p}^{+}+\vec{p}^{-}$. The $q_{T}$ variable is
the momentum component of $\vec{p}^{+}$ in the transverse plane
perpendicular to $\vec{p}_{\Lambda}$. In the case of the
$\bar{\Lambda}$ ($K_{S}^{0}$) particle one should exchange
$\vec{p}_{\Lambda}$ with $\vec{p}_{\bar{\Lambda}}$
($\vec{p}_{K_{S}^{0}}$) in the above definitions.
Fig.~\ref{fig:arm_Lambda_Kaon} is the 2-dimensional $\alpha-q_{T}$
scatter plot which shows the signatures of $\Lambda$, $\bar{\Lambda}$
and $K_{S}^{0}$ reconstructed from the experimental data
\cite{Wilrid}.

\subsubsection{$\Lambda$}

The combinatorial background is determined by rotating proton
candidate tracks around the beam axis and constructing the invariant
mass distribution. To decrease statistical errors in background
assessment, ten rotations by random angles are performed.

 \begin{figure}[tt]
  \centerline{\includegraphics[width=13.5cm,height=6.5cm]
  {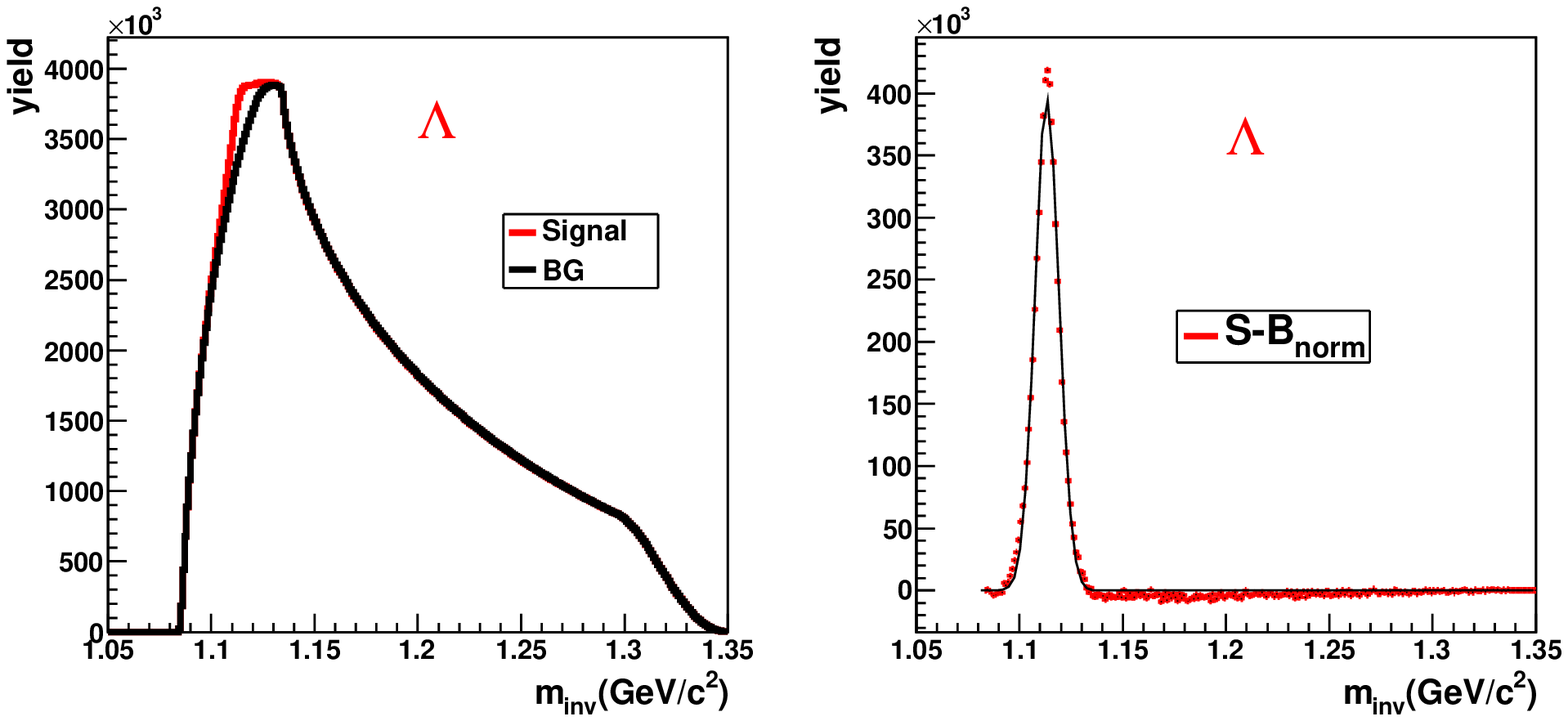}}
  \caption{Left: A small enhancement of the signal is visible 
in the region of the $\Lambda$ mass. Right: The invariant mass distribution
of the $\Lambda$ signal after subtraction of the normalized combinatorial
background is significantly non-Gaussian. \label{fig:raw_Lambda}}
 \end{figure}

The raw mass spectrum is shown in the left panel of
Fig.~\ref{fig:raw_Lambda}. The pure $\Lambda$ signal after subtraction
of the combinatorial background shown at the right side has a
non-Gaussian shape. This is owed to the facts that the observed mass
and width depend on $p_{T}$ and $y$ and that the displaced secondary
vertex is not used for recalculation of angles. The analysis is done
separately in $p_{T}-y$ windows sufficiently small in size to keep the
reconstructed $\Lambda$ mass and width practically constant. The
signal distributions are fit by a Gaussian on constant
background\footnote{found to be compatible with zero.}. 
The mass of $\Lambda$ particles strongly depends on
$p_{T}$ but practically not on rapidity, while the width depends on
both~\cite{Jovan06}. Once the mass and width are established for a
given $y$ and $p_{T}$ both are kept constant for the rest of the
analysis.

With these cuts, values of the signal-to-background ratio $S/B$ and
the significance $S/\sqrt{B}$ of about 0.04 and 500, respectively,
are obtained. Here, $S$ stands for the signal and $B$
for the background. Both quantities strongly depend on $p_T$ of the
$\Lambda$. The largest values of $S/\sqrt{B}$ reside at $p_{T} \approx
1-1.5$~GeV/c with $y\approx 2$; this is the most populated area in the
$y-p_{T}$ plane of the reconstructed $\Lambda$.

 \begin{figure}[h!]
  \centerline{\includegraphics[width=13.5cm,height=5.8cm]
  {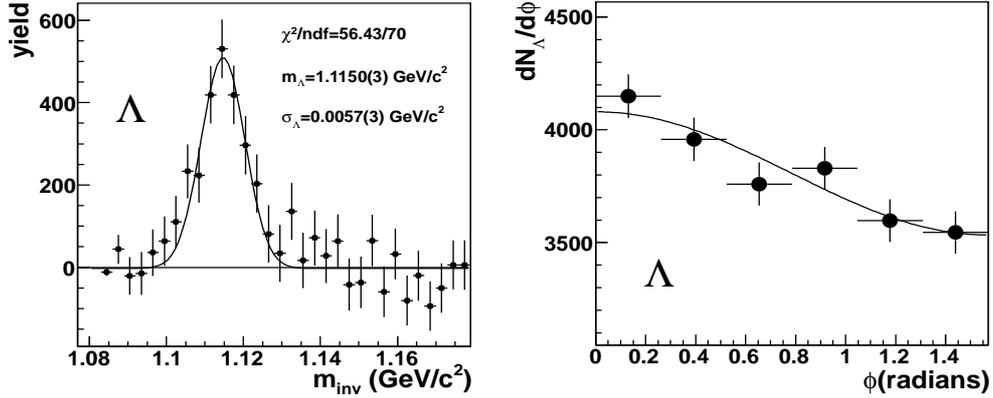}}
  \caption{Left: $\Lambda$ reconstructed for $1.62\le y \le 1.69$ in
    rapidity, $0.675\le p_{T}\le 0.8$~GeV/c in transverse momentum,
    and $15^{\circ}\le \phi \le 30^{\circ}$ in azimuth. Right:
    Elliptic flow pattern reconstructed from the $\Lambda$ yield in
    $\phi$ bins for $p_{T} \approx
    2.7$~GeV/c. \label{fig:example_Lambda}}
 \end{figure}

An example of reconstructed $\Lambda$ in a given $y$-$p_{T}$-$\phi$
bin is shown in Fig.~\ref{fig:example_Lambda} (left). The yield of
the $\Lambda$ in a given bin is obtained by fitting the invariant mass
distribution with a Gaussian.
Plotting the yield versus $\phi$ for different $p_{T}$ and
$y$ values one can construct the d$N_{\Lambda}/{\rm d}\phi$ distribution
(Fig.~\ref{fig:example_Lambda}, right). Fitting these distributions
with a function $c[1+2v_{2}'\cos(2\phi)]$, the observed elliptic flow
values $v_{2}'$ for different $p_{T}$ and $y$ were extracted. The
obtained $v_{2}'$ coefficients were corrected for the event plane
resolution as described in Sect.~\ref{flowanalysis}. 

\subsubsection{$K_{S}^{0}$}

The $K_{S}^{0}$ particles are reconstructed via the decay channel
$K_{S}^{0}\rightarrow \pi^{+}+\pi^{-}$ with branching $BR=68.95\%$ and
decay length $c\tau=~2.68$~cm \cite{PDG04}. In order to increase
statistics, the d$E$/d$x$ window is opened up to $\pm2~\sigma$ around
the nominal Bethe-Bloch energy loss value for pions.

As the $K_{S}^{0}$ particle comes from the primary vertex, a possibility 
to suppress fake track combinations is given by a cut on the radial 
distance between the point where the back-extrapolated momentum vector 
of the $K_{S}^{0}$ candidate intersects the $x-y$ plane in the primary 
vertex. The numerical value of this cut is 0.02~cm. An opening-angle cut 
$\vartheta_{\pi^{+}\pi^{-}}>$~50~mrad is applied on pair
candidates. Additionally, a cut on the $z$ position of the secondary vertex
($z>~1$~cm) was applied. In order to suppress the contamination of $\Lambda$
and $\bar{\Lambda}$ particles, an Armenteros-Podolanski cut with
$q_{T} \ge 0.12$~GeV/c was applied (see Fig.~\ref{fig:arm_Lambda_Kaon}).

 \begin{figure}[h!]
  \centerline{\includegraphics[width=13.5cm,height=6.9cm]
  {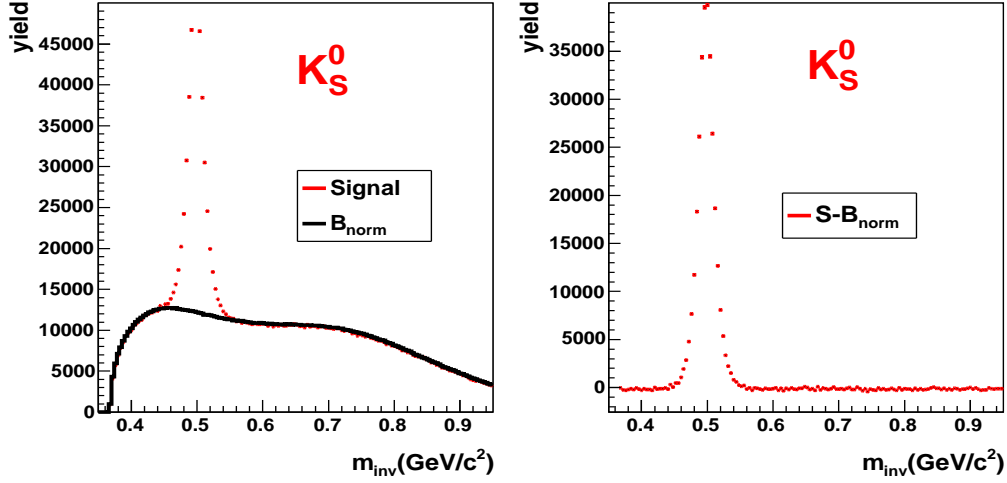}}
  \caption{Left: The invariant mass distribution of the $K_{S}^{0}$ 
    signal (red line) and the normalized combinatorial background (black
    line). Right: The invariant mass distribution of the $K_{S}^{0}$ 
    signal after subtraction of the normalized
    background. \label{fig:K_signal_thesis}}
 \end{figure}

For subtraction of the combinatorial background the mixed-event
technique is used. To preserve the event topology, only events with
similar multiplicity and orientation of the event plane are allowed
for mixing. Windows are set to $\pm$10\% and $\pm$22$^\circ$,
respectively. The event mixing is repeated 10 times.

The $K_{S}^{0}$ signal on normalized combinatorial background and 
after background subtraction is shown in Fig.~\ref{fig:K_signal_thesis}. 
Mass and width of the reconstructed $K_{S}^{0}$ exhibit $p_{T}$ and 
rapidity dependences. Values of $S/B$  and $S/\sqrt{B}$ of 
~$\approx$ 0.92 and 500 are obtained, respectively.

 \begin{figure}[h!]
  \centerline{\includegraphics[width=13.5cm,height=5.8cm] 
{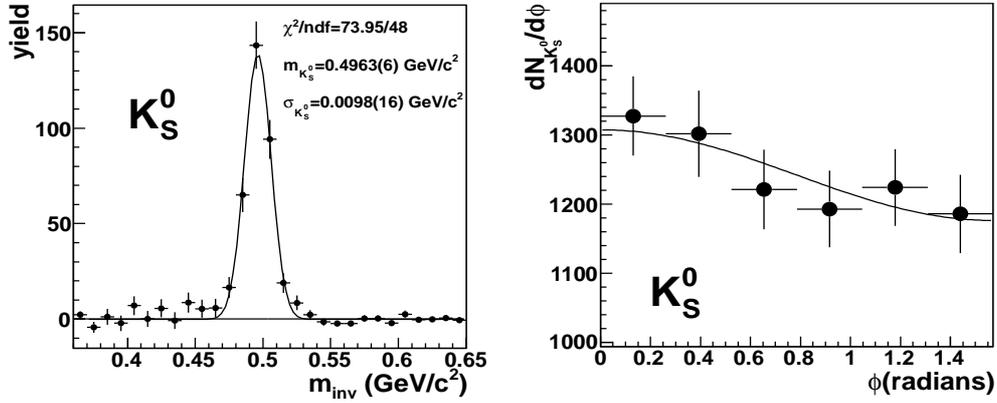}}
  \caption{Left: $K^{0}_{S}$ reconstructed for $2.075\le y \le 2.15$,
    $0.2\le p_{T}\le 0.35$~GeV/c and $45^{\circ}\le \phi \le
    60^{\circ}$. Right: Elliptic flow pattern reconstructed from the
    $K^{0}_{S}$ yield in $\phi$ bins for $p_{T} \approx
    1.7$~GeV/c. 
\label{fig:example_kaon}}
 \end{figure}

In Fig.~\ref{fig:example_kaon} an example of $K^{0}_{S}$ reconstructed
in a given $y$-$p_{T}$-$\phi$ bin (left) and a $K^{0}_{S}$ flow
pattern (right) is shown. The evaluation of the elliptic flow
magnitude is done in the same way as for the $\Lambda$
particles.

\thispagestyle{empty}
\cleardoublepage 

\section{Elliptic Flow Analysis}
\label{flowanalysis}

The flow analysis uses the event plane (EP) method, see
e.g.~\cite{PosVol,Apel98,Aga04,Jovan06} and further references
therein. We give here only a short outline to clarify notations. The
elliptic flow parameter $v_2$ is the second term in the Fourier
decomposition of azimuthal particle distributions in the plane
transverse to the beam and with respect to the orientation of the
reaction plane. However, the orientation of the reaction plane, being
not known {\it a priori}, has to be reconstructed as `event plane'
(EP) for each event,
\begin{equation}
\frac{dN}{d(\phi_i-\Phi_{EP})}=
A[~1+2v_2^{\prime}\cos(2(\phi_i-\Phi_{EP}))~].
\end{equation} 

Here, $\phi_i$ represents the azimuthal angles of outgoing particles.
The anisotropy parameter $v_2^{\prime}$ is smaller than $v_2$ in
magnitude due to the finite EP resolution.
\begin{figure}[b!]
  \centerline{\includegraphics[width=10.0cm]
  {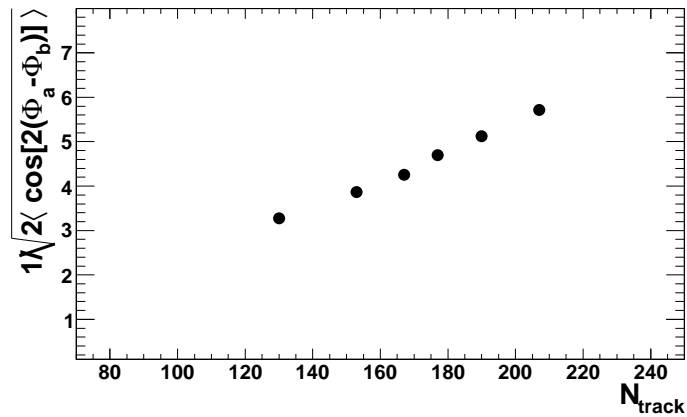}}
  \caption{The correction factor {\it vs} TPC multiplicity for the
    2-subevents method (pion data).
 \label{fig:corr}}
 \end{figure}

The azimuthal acceptance was divided into 100 adjacent slices
$a$,~$b$,~$c$,~$d$,~$a$,~$b$,~$c$,~$d$,~$a,~$...  such that every fourth slice
was assigned to a subevent $a$, $b$, $c$, or $d$, respectively. To
avoid autocorrelations, particle tracks employed for reconstruction of
the EP and of $v_2^{\prime}$ were taken from non-adjacent slices only.

Together with the reconstruction of the EP one calculates its
resolution as the average difference $<\Phi_a-\Phi_b>$ between the
EP's reconstructed from two subevents $a$, $b$. Its inverse is the
correction factor ${\cal K}$ given by
\begin{equation}
{\cal K}= \langle\mathrm{2~cos}(2(\Phi_a-\Phi_b))\rangle^{-1/2}
\label{eq:reso}
\end{equation}

by which the measured second harmonic is upcorrected,
\begin{equation}
v_2= {\cal{K}}~v_2^{\prime}.
\end{equation}

As the EP resolution depends on multiplicity, $\cal{K}$
was calculated for different centralities. Fig.~\ref{fig:corr} shows
for pion data the growing dispersion in EP orientation with multiplicity
reflecting the fact that the decrease in anisotropy wins over the
gain in statistics in deteriorating the resolution. 

To reduce autocorrelation effects, tracks chosen to be candidates for 
daughter particles were excluded from the determination of the event 
plane in case of $\Lambda$ and $K^0_S$. The event plane resolution
 \begin{figure}[tt]
  \centerline{\includegraphics[width=10.0cm] 
{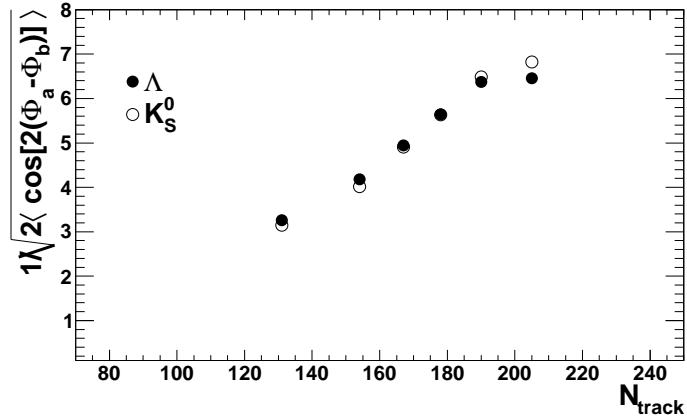}}
  \caption{Correction factors {\it vs} TPC multiplicity for $\Lambda$
(full circles) and $K^{0}_{S}$ (open circles) flow analysis.
  \label{fig:corr_Lambda_Kaon_new}}
  \end{figure}
was calculated for each multiplicity bin separately. The
correction factors ${\cal K}$ (Eq.~\ref{eq:reso}), the inverse of the 
resolution, are plotted in Fig.~\ref{fig:corr_Lambda_Kaon_new} 
over TPC multiplicity for $K^0_S$ and $\Lambda$.

Due to instrumental inhomogeneities, the reconstructed event-plane
density $<dN/d\Phi_{EP}>$ is not flat as it should be. In order to
make it flat it is enough to apply successively first the method of
recentering, and then the Fourier method of flattening~\cite{PosVol}.

\subsection{Correcting for HBT effects}
\label{hbt}

As we deal with a majority of charged pions, quantum effects among
identical bosons give rise to space-momentum correlations of
Hanbury-Brown and Twiss type (HBT): pions of the same charge tend to
cluster in azimuth if their momentum difference $|\vec{p_1}-
\vec{p_2}|$ is comparable to or below the uncertainty limit $\hbar
/R$. The critical momentum difference, with a typical source radius
$R\approx$ 5~fm, is about 40~MeV/c. Since the average pion momentum is
much larger, the effect is of short range in azimuth. The HBT effect
correlates pairs of low relative momentum; it is a positive
correlation which fakes genuine flow.

In subtracting the HBT contributions to $v_2$ we followed
Ref.~\cite{Dinh99} and used the source parameters (in the standard
Bertsch-Pratt parametrization) obtained from CERES HBT
data~\cite{adamova03,Tilsner02}. The correlation function has been
modified to take into account the effects of the Coulomb
repulsion~\cite{BaymBM96}. Several iterations of the correction
procedure were necessary to stabilize on the final value of the
integrated elliptic flow; the latter decreased thereby in relative
size by $\approx$10\%.  The results are shown in the next section.

Concerning systematic uncertainties of the HBT correction, large
relative corrections, and uncertainties are met with small magnitudes
of flow at low $p_T$; and while $v_2$ quickly increases above
0.5~GeV/c, the corrections diminish in relative size even more rapidly
and so the uncertainties. An error estimate is reached by an educated
guess based on a former study (Ref.28) where errors in the source
parameters have been included. The chaoticity parameter was given a
large error margin of $\pm$50\% in view of unknown influences of
long-lived resonances, the momentum resolution and of pairs in which
one or both pions are not primaries or of rho-decay origin. This lead
to an estimate of the relative uncertainty in the correction of
$\pm$25\%. The resulting relative errors in the HBT-corrected values
amount to $\pm$18\% and $\pm$13\% at pT= 0.25~GeV/c and 0.325~GeV/c,
respectively. At pT= 0.50~GeV/c, the systematic error is down
to $\pm$3\%.

\thispagestyle{empty}
\cleardoublepage 

\section{Postanalysis of Elliptic Flow of Candidate Pions}
\label{pionFlow}
\subsection{Overview}
\label{oview}

As illustrated in Sect.\ref{pionident}, our d$E$/d$x$ cuts
do not effectively filter out kaons and protons. In order to purify
the elliptic flow data of $\pi^-$ candidates (denoted $"\pi^-"$), the
knowledge of the $K^{-}$ fraction as a function of $p_T$ is required,
but also of $K^-$ differential elliptic flow.  We sketch the recovery
of negative-pion elliptic flow in Sect.\ref{piminus}.

Positive pions are mixed with positive kaons and protons for $p\ge$
1.2~GeV/c. In order to isolate the {\it proton} elliptic flow from the
measured $v_2$ data of $\pi^+$ candidates, $v^{``\pi^{+}``}_{2}$, we
will use the particle ratios at 158~$A$GeV which recently became
available from measurements of Pb+Pb collisions by the NA49
Collaboration~\cite{NA4908}.

The composition of the particle flux as accepted by the CERES
spectrometer and filtered by the previous analysis cuts~\cite{Jovan06}
has to be reconstructed. It is mandatory to properly simulate the
effects of the pion-tuned d$E$/d$x$ cut on kaons and protons. The
d$E$/d$x$ filter requires knowledge of particle momenta $p$=
$p_T$/sin\,$\vartheta$.  Unfortunately, at the time this analysis was
started, the information on polar angle $\vartheta$ from the doublet
of Silicon-Drift Detectors was no longer accessible.  Without it, the
resulting spread in sin$\vartheta$ over the acceptance, of almost a
factor of two, would have blurred the d$E$/d$x$ resolution. To avoid
such degradation in quality, recourse was taken to a full Monte-Carlo
(MC) simulation.  We give details on the reconstruction of proton
$v_2$ in Sect.\ref{v2protons}.

In a first approach to the acceptance correction we used analytic
methods; although these calculations were made obsolete by the MC
simulation, they serve as a valuable check on the final
result. Besides, some are quite instructive and have been instrumental
for preparing the input parameters of the MC simulations.

\subsection{Processing elliptic flow of $\pi^-$ candidates}
\label{piminus}

The expression linking the $\pi^-$ differential elliptic flow 
$v^{\pi^{-}}_{2}$ to the available data is
\begin{equation}
v^{\pi^{-}}_{2} = 
v^{``\pi^{-}``}_{2} + 
r_K^{-} ~(v^{``\pi^{-}``}_{2} - v^{K^{-}}_{2}).
\label{eq:v2piminus}
\end{equation}
\vspace{-0.4cm}

Here, $v^{K^{-}}_{2}$ is the flow parameter of $K^-$, $r_K^-$ denotes
the particle ratio $r_K^-= N_{K^{-}}/N_{\pi^{-}}.$ All quantities are
functions of $p_T$.  Statistical errors enter in a way that forbids to
apply standard error propagation. Therefore, a simulation was
performed treating the three experimental inputs as Gaussian random
variables with widths equal to their statistical errors.  For every
channel of the $p_T$ spectrum to be incremented, the flow parameter
$v_2(p_T)^{\pi^{-}}$ is given by the average over many trials, its
statistical error by the dispersion.

As input for $v^{K^{-}}_{2}$ we use the differential elliptic flow
data for $K_S^0$ which will be presented in
Sect.\ref{strangeFlow}. Because of the very similar mass and quark
contents of kaons, a possible difference in $v_2$ can be considered
small compared to the present accuracy (see below).

 \begin{figure}[tt]
  \centerline{\includegraphics[width=6.5cm] 
{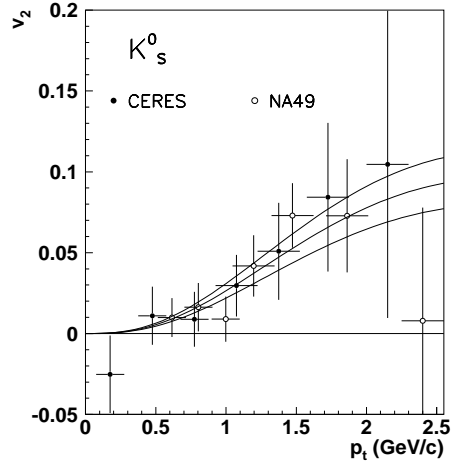}}
  \caption{$K_S^0$ differential elliptic flow combining CERES data
(this work; filled symbols) with NA49 data (ref.\cite{Blume08}; open
symbols). Centralities are (5.3~-~13)\% of $\sigma_{geo}$ for CERES
(average 9.8\%) and top 13\% for NA49 data.  The three lines show the
best-fit within the 1-$\sigma$ errors bands. Errors are statistical.
  \label{fig:v2-K0S-combined}}
  \end{figure}

To improve on the statistical significance of the $K_S^0$
elliptic flow data, we have combined our results with those of
NA49~\cite{Blume08,Stefanek06}. The latter data have been collected at the
top 13\% of $\sigma_{geo}$.  The two data sets are shown in
Fig.~\ref{fig:v2-K0S-combined} together with a 1-parameter fit
according to $v_2(p_T)= A~p_T^3$~exp$(-p_T)$. The best fit resulted
in $\chi^2/ndf$=~0.44. Shown is the best-fit curve for $A=
7.18\cdot10^{-2}$ in the centre, sandwiched between the 
$\pm$ 1$\sigma$ statistical error bands of $\pm$ 17\% relative.

To quantify the unknown admixture of $ K^-$, we have used the $K^-$
and $\pi^-$ transverse momentum spectra measured by the NA49
collaboration~\cite{NA4908,NA4905}. Details are given in the
following section for the analogous case of positive-pion
candidates. All steps and procedures, like centrality matching, d$E$/$dx$
cut and acceptance corrections within the Monte-Carlo simulation,
fully apply also to the negative-pion sample. 

\subsection{{Processing elliptic flow of $\pi^+$ candidates}
\label{v2protons}}
\subsubsection {Outline}

To determine the differential elliptic flow of the minor component
of protons from measured $v_2(p_T)$ data of $\pi^+$ candidates, we
make the simplifying assumption $v^{\pi^{+}}_{2}\approx
v^{\pi^{-}}_{2}$, and use the differential flow of $\pi^-$ derived in
Sect.\ref{piminus} as substitute for $v_2(\pi^+)$ in the proton flow
analysis below. For a discussion of possible violations of this
assumption we refer to Sect.\ref{digression} and for our assessment of
related uncertainties to Sect.\ref{bound}.

The measured elliptic flow $v^{``\pi^{+}``}_{2}$ of the $"\pi^+"$
candidate sample contains a contribution of proton elliptic flow
$v^{p}_2$,
\begin{equation}
\label{eq:PiPlusSample}
v^{``\pi^{+}``}_{2} = (N_{\pi^{+}}~v^{\pi^{-}}_{2}+ N_{K^{+}}~v^{K^{+}}_{2}+ 
N_{p}~v^{p}_{2})/(N_{\pi^{+}}+N_{K^{+}}+N_{p}).
\end{equation}
\vspace{-0.5cm}

More explicitly, the unknown magnitude $v^{p}_2$ of the proton elliptic
flow is derived as
\begin{equation}
\label{eq:KProtonFlow}
v^{p}_{2} = 
((1+ r_{K^+}+ r_p)~v^{``\pi^{+}``}_{2} 
   - v^{\pi^{-}}_{2} - r_{K^+}~v^{K^{+}}_{2})/r_p.
\end{equation}
\vspace{-0.5cm}

On the r.h.s., we substitute the measured $v_2(K^{0}_{S})$ for
$v_2(K^+)$ . The only quantities yet unknown are the particle ratios 
$r_K^{+}= N_{K^{+}}/N_{\pi^{+}}$ and $r_p= N_{p}/N_{\pi^{+}}$ specifying the
contents of $K^{+}$ and protons in the $``\pi^{+}``$ sample, respectively.  

\subsubsection{Differential particle yields}

\begin{figure}[t]
  \centerline{\includegraphics[width=7.5cm]{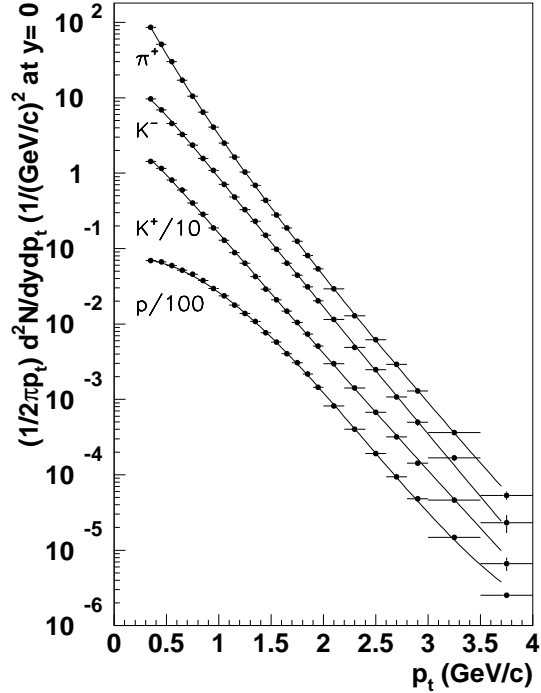}}
  \caption{Invariant $p_T$-differential yields of $\pi^+$, $K^+$,
$K^-$ and protons from Ref.~\cite{NA4908} with statistical errors,
shown together with 3-parameter fits obtained by minimizing
$\chi^2$. $K^+$ data and curve are divided by 10, proton data by
100. The $\pi^-$ spectrum, very close to that of $\pi^+$, is left out
in order not to overload the figure.
  \label{fig:dndpt}}
  \end{figure}

Invariant yields of charged pions, charged kaons and protons at midrapidity
for inelastic Pb-Pb collisions at $\sqrt {s_{NN}}= 17.3$~GeV,
\begin{equation}
\label{eq:ptspectra}
\left.\frac{{\rm d}^2N(p_T)}{{\rm d}y{\rm d}p_T}=
\frac{1}{2\pi p_T}~\frac{{\rm d}^2N}{{\rm d}y{\rm d}p_T}\right\vert_{y= 0},
\end{equation}
which are relevant for the present study are those displayed in
Fig.~8 of Ref.\,\cite{NA4908}. The ~$p_T$~ spectra are adjusted to
CERES centrality as described below.  For the purpose of sampling
$p_T$ values at random which obey the proper density distributions,
spectra are fit by 3-parameter exponential functions.  The fits to the
data are shown in Fig.~\ref{fig:dndpt}. The authors of
Ref.\,\cite{NA4908} have estimated the systematic errors,
compounded from d$E$/d$x$, feed-down yields and acceptance corrections,
as 2.2\% for charged pions and $K^-$, 4.5\% for $K^+$, and 3.7\% for
protons. The fits also shown in Fig.~\ref{fig:dndpt} deviate form the
data points by typically less than $1\%$.

\subsubsection {Matching Centralities}

Our mid-central collision window (5.3-14.5)\% of $\sigma_{geo}$ does
not find a close match among NA49 centrality classes. Since data for
the closest (5-12.5)\% selection were not available~\cite{Blume09}, we
have used a linear combination of centrality classes (0-5)\% and
(12.5-23.5)\%. By visual inspection of the NA49 $p_T$ spectra for
different centrality classes, a composition with
equal weights seemed most appropriate. We also calculated the weighted
means $\langle\sigma/\sigma_{geo}\rangle$ for the two slices
representing the NA49 centrality classes 1 and 3 to determine the
required composition quantitatively: the mixture of 55\% of class-1 centrality
(0-5)\% combined with 45\% of class-3 centrality (12.5-23.5)\% well
reproduces the mean centrality of 9.8\% of CERES mid-central triggers
and was used.

\subsubsection {Acceptance}

\begin{figure}[t]
  \centerline{\includegraphics[width=5.5cm]{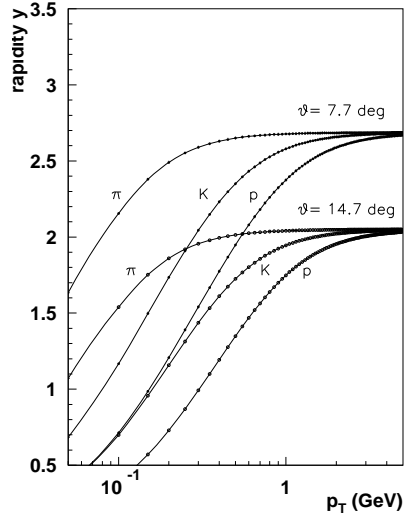}}
  \caption{The boundaries of CERES acceptance defined by polar
angles $\vartheta$= 7.7$^\circ$ and 14.7$^\circ$ shift
 away from central rapidity ($y_{mid}=\,2.91$) with decreasing $p_T$,
the stronger the more massive the particle species, as shown here for
pions, kaons, and protons. The $\eta$-acceptance 2.05\,$<\eta<\,$2.70 
coincides with y-acceptance at infinite momentum ($\beta$=~1).
  \label{fig:getrap}}
  \end{figure}

CERES accepts a cone in polar angle of
$7.7^\circ<\vartheta<14.7^\circ$. The small acceptance in
pseudo-rapidity, $\Delta\eta$\,=\,0.65 units, spans the range
2.05\,$<\eta<$~2.70 and is close to midrapidity, $y_{mid}$\,=
2.91. For $p_Tc$ smaller than particle mass $mc^2$, the acceptance is
shifted down in rapidity with decreasing $p_T$ as shown in
Fig.~\ref{fig:getrap}.

The practicable assumption is made that the doubly-differential yields
of Eq.~\ref{eq:ptspectra} factorize in $y$ and $p_T$. Particle
yields within the CERES acceptance are then obtained by integrating the
corresponding rapidity distributions between the $p_T$-dependent rapidity
corners denoting the acceptance for given $m$ and $p_T$, i.e. those
shown in Fig.~\ref{fig:getrap}. We use the NA49 parametrization
of the rapidity distributions by two identical Gaussians of width
$\sigma$ which are shifted from mid-rapidity by equal and opposite
amounts $\pm y_S$~\cite{Afanasiev02},
\begin{equation}
\label{eq:rapdensity}
\frac{dN}{dy}= 
{\cal N}\left[exp(- \frac{(y- y_S)^2}{2\sigma^2})+ 
           exp(- \frac{(y+ y_S)^2}{2\sigma^2})\right].
\end{equation}
 \begin{table}[b]
  \centerline{\includegraphics[width=6.5cm]{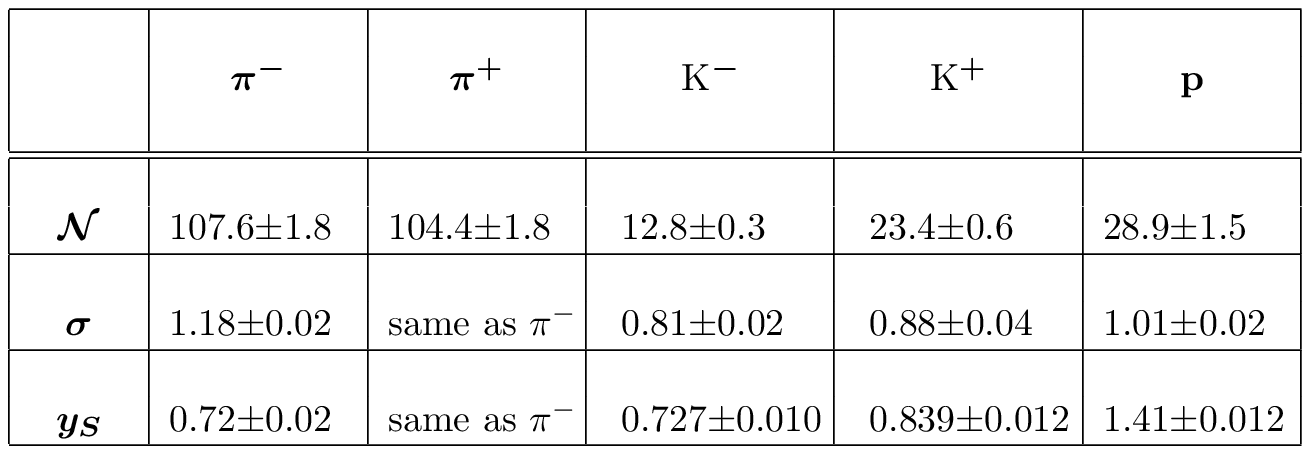}}
  \caption{Parameters used to simulate d$N$/d$y$ distributions using 
   Eq.~\ref{eq:rapdensity}. From Ref.~\cite{Afanasiev02}. See text.
  \label{fig:ypar}}
  \end{table}
The values used in the present calculation for parameters ${\cal N},
\sigma$, and $y_S$ are listed in Table~\ref{fig:ypar}. Those for
charged pions and kaons were taken from Table~III of
Ref.~\cite{Afanasiev02}.  The yield parameter for $\pi^+$ was
downscaled relative to that for $\pi^-$ by the same factor by which
both the total multiplicities and the values of $\langle {\rm d}N/{\rm
d}y\rangle_{|y|<0.6}$ are observed to scale, by inspection of Table~II
of Ref.~\cite{Afanasiev02}. For protons, the d$N$/d$y$ parameters were
determined by best-reproducing the distribution labeled CC2 in
Fig.\,6.6 of Ref.~\cite{Utvic08}. This centrality class corresponds to
(5-12.5)\% of $\sigma_{geo}$, close to CERES centrality.

Particle invariant yields have to be averaged over the
acceptance before taking ratios; those will no longer reflect yields
at mid-rapidity alone, but will also depend on the shapes of the
d$N$/d$y$ distributions with appreciable differences among the particle
species. In addition, the shift in rapidity is taken into account.

A suitable reference for the acceptance corrections of the invariant
particle yields is the geometrical acceptance for massless particles
($\beta$=\,1) for which
 \begin{figure}[t!]
  \centerline{\includegraphics[width=6.5cm]{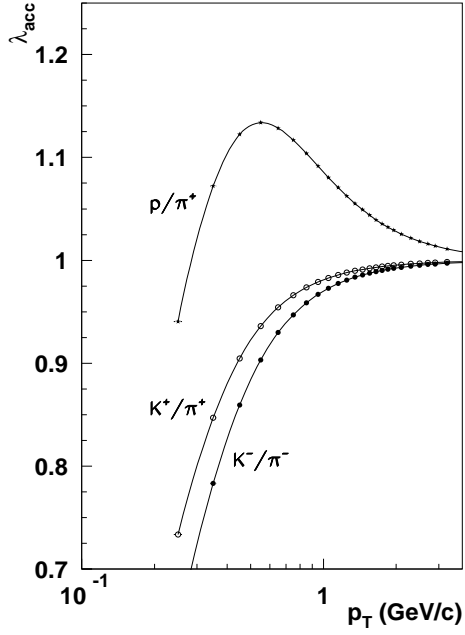}}
  \caption{Acceptance correction functions $\lambda_{acc}$ 
of Eq.~\ref{eq:acc} transforming NA49 particle ratios to CERES acceptance.}
  \label{fig:acc}
  \end{figure}
pseudo-rapidity coincides with rapidity,
$\eta\equiv y$; Fig.~\ref{fig:getrap} would consist of two
horizontal straight lines at $y= -0.86$ and at $y= -0.21$\footnote{we
quote centre-of-mass rapidities downshifted from lab rapidities by the
rapidity of the centre of mass in the lab, 2.91.}. With such
reference, the acceptance correction is quantified by the ratio
 \begin{equation}
\lambda_{acc}(p_T,\,m)= \left[~~\int_{y(\eta_{\,low};~m,\,p_T)}^
    {y(\eta_{\,high};~m,\,p_T)}(\frac{{\rm d}N}{{\rm d}y})\,{\rm d}y~\right]~/~
    \left[~~\int_{y=\,\eta_{\,low}=\,-0.86}^{y=\,\eta_{\,high}=\,-0.21} 
     (\frac{{\rm d}N}{{\rm d}y})\,{\rm d}y\right].
\label{eq:acc}
\end{equation}
The resulting correction functions are shown in Fig.~\ref{fig:acc} for
the three particle ratios. The rapidity shift has opposite effects on
the yield of protons and kaons: due to the rather compact distribution
of kaons compared to the well-separated double humps in the d$N$/d$y$
distribution of protons, the particle composition around 0.5 GeV/c is
reduced by about 6\% in kaons but enriched by 15\% in protons over
the $\beta=1$ reference.\\

\subsubsection {Monte-Carlo simulation}

\noindent
Transverse momenta for given particle type are generated from the
respective probability densities, using standard
methods~\cite{numrec92}. Before doing so, $p_T$ spectra are adjusted
to CERES centrality as described.  `Events' are filtered by conditions
of acceptance and d$E$/d$x$ cut. Once $p_T$ is chosen, boundaries of the
acceptance for the respective particle species are defined in
rapidity, as shown in Fig.~\ref{fig:getrap}. Then $y$ is chosen at
random from the density distribution of Eq.~\ref{eq:rapdensity} using
the parameters of Table~\ref{fig:ypar}.  Rapidity values outside the
acceptance window result in rejection of the event.
\begin{table}[b]
\begin{tabular}{ | c | c | c | c |}  
         \hline
    {particle ratio}&{K$^+/\pi^+$} &{K$^-/\pi^-$}& {p/$\pi^+$} \\
         \hline
      {\makebox[1cm][c]{$x_S$ ratio}}& {\makebox[2.4cm][c]{1.401~$\pm$0.026}} & 
{\makebox[2.4cm][c]{1.473$~\pm$0.015}}&
{\makebox[2.4cm][c]{1.415~$\pm$0.016}}\\ 
         \hline
\end{tabular}\\
  \caption{Factors transforming MC-generated yield ratios at midrapidity
  to ratios of yields averaged over CERES' y-acceptance
  (Eq.~\ref{eq:xS}). Errors are propagated from Table~\ref{fig:ypar}.
  \label{fig:xSpar}}
  \end{table}
At this stage, the entire event is defined: the particle momentum $p$
is calculated from $m$, $p_T$, and $y$, which also fixes the
Bethe-Bloch most-probable d$E$/d$x$ value. Gaussian noise is added to
simulate the experimental resolution $\sigma$=~0.10~d$E$/d$x|_{BB}$.  The
historical filter set to (0.85-1.15)d$E$/d$x|_{BB}$ of {\it pions}
during data analysis is activated. Now, the survival fractions of
kaons and protons are determined as error integrals over the d$E$/d$x$
distributions between the $p_T$-dependent boundaries of the cuts on
charged pions. The $p_T$ spectra are incremented by survival fractions
$f$, $0<f<1$.

Monte-Carlo generated spectra require subsequent normalization to the
experimental data after which they were modelled\footnote {Since only
particle {\it ratios} enter into Eq.~\ref{eq:KProtonFlow}, there is
allowance for one free parameter common to all spectra.}.  The NA49
particle yields are summed up from 0.30~GeV/c, the lower boundary of
the first entry in the published spectra, to a reasonable upper
boundary, i.e. to 4.0~GeV/c. The normalization is achieved by setting
the initial MC-yield, prior to any filtering and summed-up over the
said $p_T$ range, equal to the sum over the respective data spectrum.

The $p_T$-integrated yield is multiplied by a factor $x_S$ to transform 
the mid-rapidity yield to the yield averaged over acceptance,
\begin{equation}
\label{eq:xS}
x_S= 
\langle {\rm d}N/{\rm d}y \rangle_y^{\rm CERES}~/~{\rm d}N/{\rm d}y|_{y=0}~.
\end{equation}
The transformation factors for particle ratios are listed in
Table~\ref{fig:xSpar}. Using identical binning, the statistical data
errors are adopted.

\thispagestyle{empty}
\cleardoublepage 

\section{Elliptic flow of charged pions: Results}
\label{piminusFlow}

We present here the elliptic flow results of negative pions,
illustrating the effect of the HBT correction and the subtraction of
the $K^{-}$ component as outlined in Sect.~\ref{hbt}. 

\begin{figure}[ht]
  \centerline{\includegraphics[width=10.5cm]
  {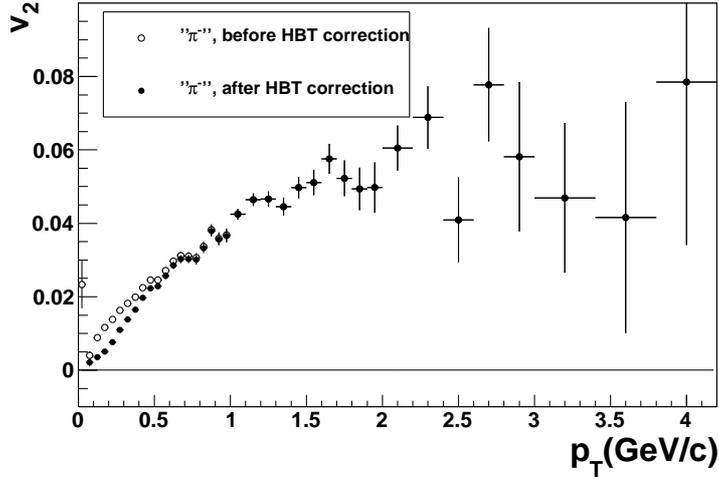}}
\caption{Elliptic flow $v_2(p_T)$ of $\pi^-$ candidates
before (open circles) and after (filled circles) correction for 
the HBT effect (mid-central trigger).
 \label{fig:preposthbt}}
 \end{figure}

The elliptic flow of negative pion candidates is shown in
Fig.~\ref{fig:preposthbt} before and after the HBT correction.  It is
quite satisfactory to see that data start very close to zero and
follow a quadratic $p_T$ dependence for small $p_T$.

\begin{figure}[bb]
  \centerline{\includegraphics[width=6.0cm] 
{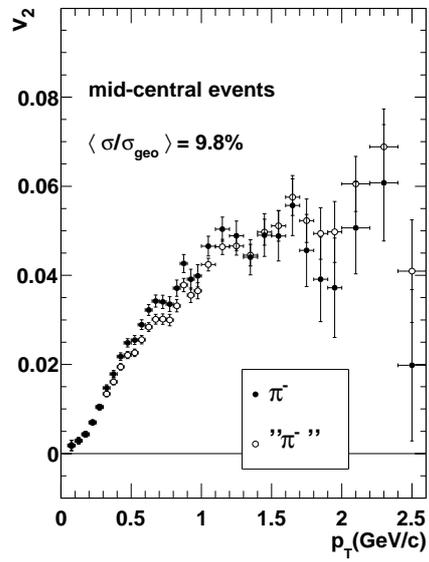}}
  \caption{Differential elliptic flow $v_2(p_T)$ of negative pion
   candidates (open circles); and of negative pions after 
   removing the $K^-$ admixture (filled circles). The $v_2$ spectra 
  are corrected for HBT correlations. Statistical errors 
  are compounded from all sources entering Eq.~\ref{eq:v2piminus}
(mid-central triggers).
  \label{fig:PiMinusCorrect}}
  \end{figure}

The $p_T$-differential elliptic flow $v_2(p_T)$ of negative pion
candidates is displayed in Fig.~\ref{fig:PiMinusCorrect}, together
with the $v_2(p_T)$ spectrum of negative pions obtained by subtracting
the $K^-$ component. Both spectra are corrected for HBT correlations
as outlined in the previous section. Statistical errors are obtained
by Monte-Carlo sampling along Eq.~\ref{eq:v2piminus}, treating all
input data as Gaussian random variables.  The estimate of 25\% for the
relative systematic error in the HBT correction amounts to about 15\%
of the corrected value around $p_T$= 0.30~GeV/c (for details see
Sect.~\ref{hbt}).

The $K^-$ impurity is seen to have only a minor effect. The correction
starts out positive, raises the $v_2$ value by about 0.005 at
0.6~GeV/c, decreases in size and changes sign at $p_T\approx$
1.4~GeV/c; it turns negative where $v_2$ of pions and kaons cross each
other.  The smallness of the correction follows from the rather small
$N_{K^{-}}/N_{\pi^{-}}$ ratio and the two elliptic flow
parameters being close in magnitude.

\thispagestyle{empty}
\cleardoublepage 

\section{Elliptic flow of protons}
\label{protonFlow}

\subsection{Systematic uncertainties}
\label{sys}

Proton $v_2$ parameters in the present study are obtained as
differences between two `large numbers', the two pion flow
measurements entering Eq.~\ref{eq:KProtonFlow}. Alarming as this might
be in anticipation of relative errors getting out of control, the
following evaluation will dispel such concern. A decisive advantage is
that both measurements were performed and analyzed in identical
settings.\footnote{An exception is the different sign of track
curvature in the TPC. The zero-deflection was fine-tuned by
reconstructing the same mass for $\Lambda$ and $\bar{\Lambda}$
~\cite{NIM08}.}  In addition, CERES' full azimuthal coverage minimizes
uncertainties in the EP determination and the flow measurement itself.

Other uncertainties common to pions of both charges are related
to the centrality determination, and the $dE/dx$ and acceptance
cuts. Summing up the individual error estimates in quadrature yields
6\%.  A `common-mode' error of this size is applied to both pion flow
parameters and the response is calculated for the full $p_T$ range of
interest. The deviations in proton $v_2$ rise from negligible levels
with increasing $p_T$ to $\Delta v_2\approx$~0.008 at
2.5~GeV/c.

Uncertainties linked to HBT correlations and the corrective measures
taken are common to both pion flows and treated as {\em relative}
errors. The HBT correction is largest at the lowest $p_T$ bin of
0.325~GeV/c.  A systematic relative uncertainty of 15\% in the
corrected values at $p_T\approx$~0.30~GeV/c causes an uncertainty in
proton flow of $\Delta v_2\leq$~0.004 which decreases quickly with
rising $p_T$ to practically vanish around 0.6~GeV/c.  The directly
identified protons (the four asterisk points in Fig.~\ref{fig:v2prot}
below) are free of HBT correlations and related uncertainties.

One may think also of uncertainties in {\rm only one} of the two pion
flow parameters which remain unbalanced. The subtraction of the $K^-$
component from the $"\pi^-"$ candidate flow parameter is the only such
example we are aware of. An estimated relative error of 8\% in $v_2$
of $K^-$ combined with a 5\% relative uncertainty in the $K^-/\pi^-$
particle ratio has little impact.

We have not corrected the pion elliptic flow data for non-flow
correlations apart from HBT. Autocorrelation effects of short range
between the samples used for flow measurement and for determining 
the event plane, respectively, were avoided by accepting only
non-contiguous combinations. In the proton flow analysis we may assume
that on average non-flow correlations in the positive and the negative
pion samples cancel each other; for the remaining proton elliptic flow
we argue that jet-like correlations, the main physics source of
non-flow correlations for pions, have negligible proton content at
this energy.

So much to systematic uncertainties in CERES data. The MC simulation
introduces additional uncertainties.  Errors in particle ratios have
been derived from d$n$/d$p_T$ spectra of Ref.\cite{NA4908} as 5.0\%
for $K^+/\pi^+$, and 4.3\% for $p^+/\pi^+$.  The MC-generated spectra
carry small normalization errors, quoted in
Table~\ref{fig:xSpar}. Further errors arise by simulating $dE/dx$ cuts
and acceptance corrections, to our estimate of 3\% each; they add up
to 6.8\% and 6.1\% for the $K^+/\pi+$ and $p/\pi+$ ratios,
respectively. The calculated uncertainties induced in proton $v_2$ are
largest at low $p_T$ but remain below 0.003.  More harmful is the
relative uncertainty in $v_2$ of $K^0_S$: although deviations in
proton $v_2$ start well below 0.001, they approach 0.008 at the
largest $p_T$.

\subsubsection{A digression: pion flow asymmetry}
\label{digression}

Our derivation of proton $v_2(p_T)$ rested on the plausible assumption
that $v_2(\pi^+)$ and $v_2(\pi^-)$ are sufficiently close.  With
recent dicussions of exciting new physics related to chiral magnetic
effects~\cite{burnier11}, an asymmetry in elliptic flow of particles
and antiparticles has come into focus and its observation was recently
reported by the STAR Collaboration~\cite{bes11}. Preliminary data for
minimum-bias Au+Au collisions at $\sqrt {s_{NN}}$= 7.7 and 11.5~GeV show
sizeable asymmetries\footnote{we changed the sign so that ${\cal A}>0$ 
in accord with the STAR results.} 
\begin{equation}
\mbox{${\cal A}= (v_2(\pi^-)- v_2(\pi^+))/v_2(\pi^+)$}
\end{equation}
 at very low $p_T$.\footnote{we read from the
J.~Phys. paper~\cite{bes11}, Fig.~2 (top right) ${\cal A}$ values for
$\sqrt {s_{NN}}$= 7.7, 11.5, and 39 GeV and $p_T$= 0.3~GeV of 30\%,
26\%, and 10\%, respectively, which at $p_T$= 0.5~GeV, reduce to to
11\%, 8\%, and 4\%; minimum-bias trigger.}

There are reasons to assume the pion flow asymmetries will be
considerably reduced towards more central collisions. We will go along
some of the arguments that have been suggested, mostly by the authors
of Refs.~\cite{burnier11,bes11} with emphasis on the expected $p_T$
and centrality dependence. We start with the more conventional scenarios.

Resonance decays have been discussed for effects on pion elliptic
flow~\cite{grecoko05} and possible violations of the naive quark
coalescence model~\cite{molvol03}. Other than $\rho$-mesons,
$\Delta(1232)$-resonances may induce charge-dependent effects in flow
by two reasons: (i) the isospin asymmetry between $u$ and $d$
flavours\footnote{the neutron-to-proton ratio for Pb-Au is 1.52.}
gives an edge to excitation of $\Delta^\circ$ and $\Delta^-$ over
$\Delta^+$ and $\Delta^{++}$, causing a surplus of $\pi^-$ over
$\pi^+$ decays. (ii) As $\Delta$'s are likely to be recombined from
two hadrons ($N\pi$), their $v_2$ will increase over strict
number-of-constituent-quark (NCQ) scaling, and decay pions are to show
{\it larger} $v_2$ than directly produced
pions~\cite{grecoko05,nonaka04}. Since $\Delta$-decay pions are very
soft, the flow asymmetry should show at very low $p_T$.

The asymmetry is transmitted only by decay pions which leave the fireball
without being rescattered prior to thermal freeze-out. The condition
is met by only a small fraction of $\Delta$'s decaying close to
thermal freeze-out.\footnote{reduction of resonance yields relative to
stable particles was described recently~\cite{markert07}.} The
number of unscathed decay pions should be rather independent of
centrality: production of $\Delta$'s grows with $N_{ch}$ as do
rescattering losses due to increased density. However, the asymmetry
in the pion sample for more central collisions is reduced by an
increasing share of thermal pions ($\propto N_{ch}$). At the bottom
line, ${\cal A}$ induced by resonance decays should scale inversely
with $<N_{ch}>$. For our 10\% trigger we estimate a drop to about 40\%
relative to the \mbox{(0-80)\%} minimum-bias trigger of the STAR data.

Another scheme for violation of NCQ scaling has been
proposed~\cite{dunlop11} in which partons carry larger amounts of flow
strength arriving at midrapidity from stopped baryons, than those from
$q\bar{q}$ pairs. This effect yields the correct sign, but is rather
smallish: at $\sqrt{s_{NN}}$= 8.86~GeV, the asymmetry is about 1\%. At
17.3~GeV, it is further reduced by a factor 2.6\footnote{using the
parameters of Ref.~\cite{dunlop11} and particle data from NA49 at
17.3~GeV.}. In this model, the fraction of constituent {\it u} quarks
transported by baryon stopping decreases from 0.50 to 0.24 between
8.86~GeV and 17.3 GeV, respectively.\footnote{quoting values of $X_T$
defined in Ref.\cite{dunlop11}.} This suggests an equal reduction in
proton density at midrapidity and hence of Coulomb repulsion of
positive pions as a possible cause of the asymmetry, whatever its
importance might be.

The Chiral Magnetic Effect (CME): a transient magnetic field induces
an electric current at finite baryo-chemical potential in presence of
a chiral asymmetry between left- and right-handed quarks. The current
generates an electric quadrupole with positively charged poles `above'
and `below' the reaction plane causing the asymmetry $v_2(\pi^-)>
v_2(\pi^+)$~\cite{burnier11}. The numerical estimates
of~\cite{burnier11} are by and large consistent with the preliminary
data~\cite{bes11}. It is seen from Ref.\cite{kharzeev08} that
observables which measure the correlation of positive/negative charges
to the reaction plane~\cite{voloshin04} are reduced by one order of
magnitude between semi-peripheral collisions of 50-60\% centrality and
mid-central collisions below 20\% of the geometrical cross
section.\footnote{referring to the plot of $a_{++},a_{--}$ {\it vs}
centrality in Fig.~3 of Ref.\cite{kharzeev08}.} It is plausible to
assume that a CME-inflicted pion flow asymmetry is reduced also by one
order of magnitude at our 10\% centrality, compared to the minimum
bias results.

In summary, $\Delta$-resonance decays and the CME seem plausible
candidates for inducing an isospin asymmetry on pion flow at low
$p_T$, of the correct sign and possibly of the magnitude indicated by
the preliminary STAR data~\cite{bes11}.  The arguments presented suggest that
${\cal A}$ is considerably reduced in more central collisions compared to
minimum-bias collisions studied by STAR.

\begin{figure}[tt]
  \centerline{\includegraphics[width=5.5cm] 
{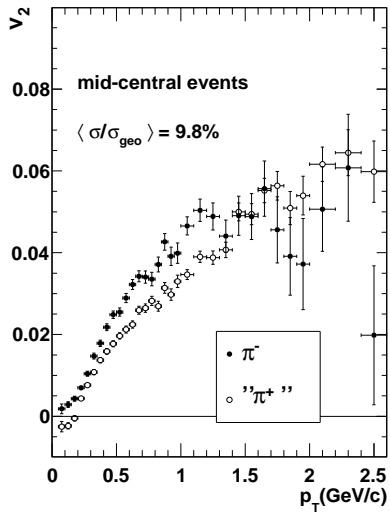}}
  \caption{Differential elliptic flow $v_2(p_T)$ of $\pi^+$ candidates
   (open circles) and of identified negative pions (filled circles)
   shown already in Fig.~\ref{fig:PiMinusCorrect}. The two flow
   spectra differ by the $K^+$ and proton admixtures to the former,
   open-circled spectrum. A possible pion-flow asymmetry has been neglected.
  \label{fig:PiPlus}}
  \end{figure}

\subsection{Results}
\label{res_pflo}

The primary data entering the proton flow analysis are displayed in
Fig.~\ref{fig:PiPlus}. The indication of a negative excursion in $v_2$
of the $"\pi^+"$ candidates at low $p_T$, absent in $\pi^-$ flow, gave
the incentive for the present reanalysis.  The derivation of proton
$v_2(p_T)$ is
\begin{figure}[tt]
  \begin{minipage}{1.0\linewidth} 
  \centerline{\includegraphics[width=6.5cm]
      {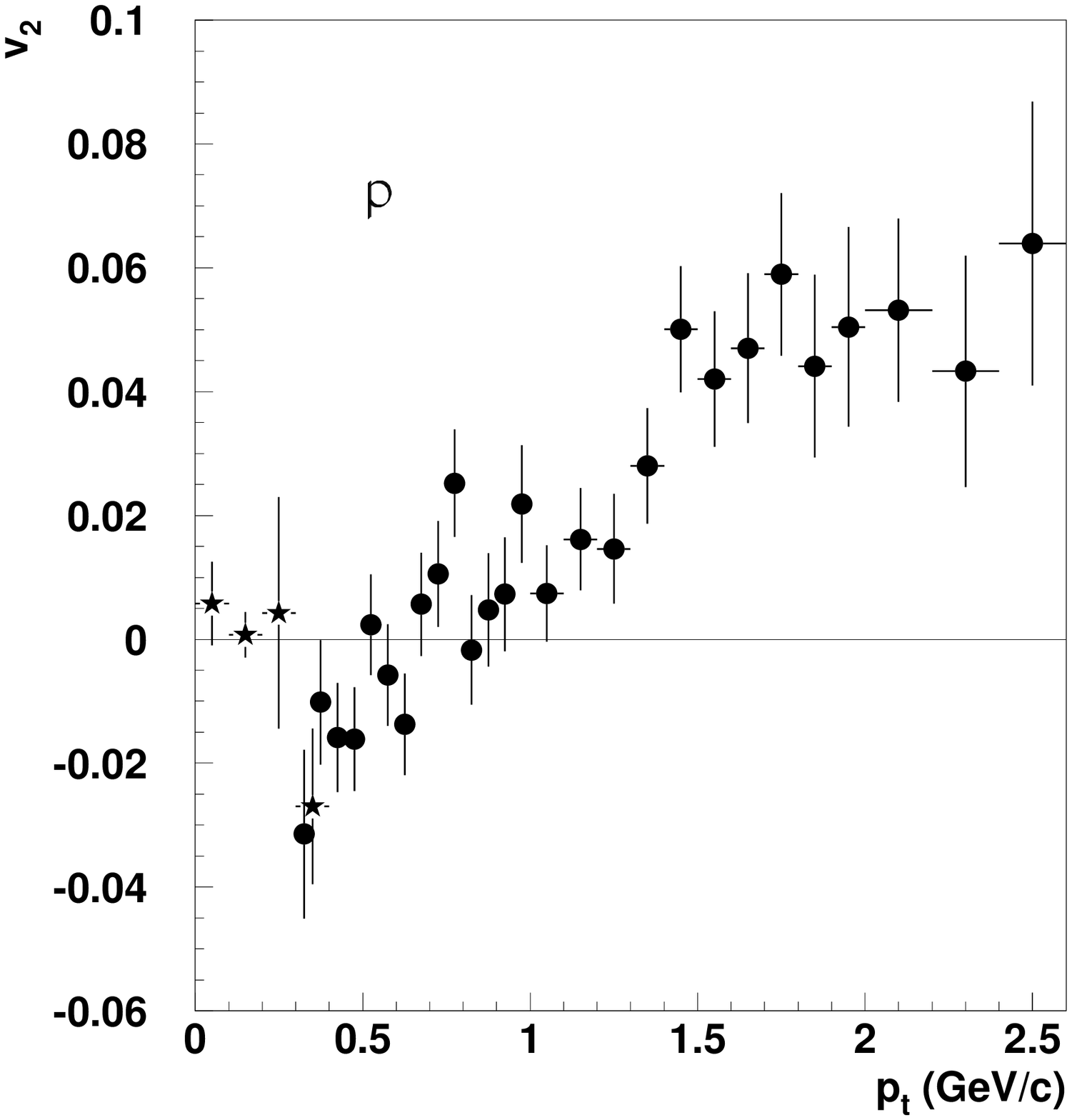}}
  \end{minipage} 
\vspace{-0.4cm}

  \begin{minipage}{1.0\linewidth} 
  \centerline{\includegraphics[width=6.5cm]{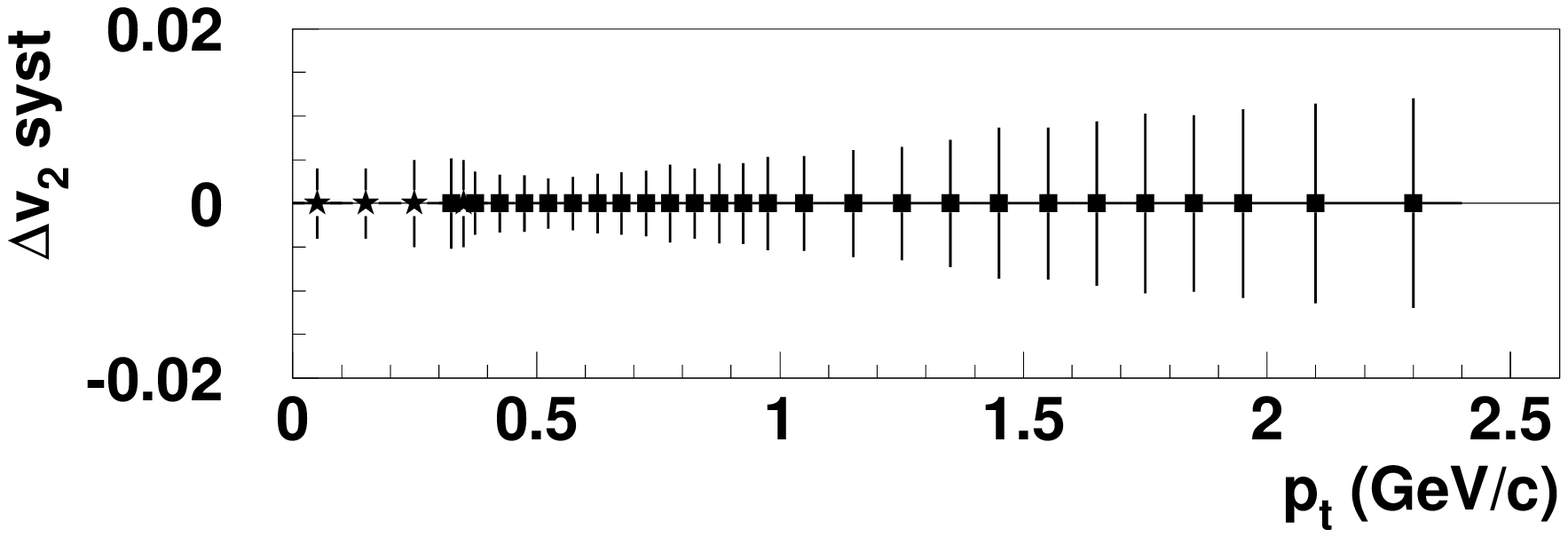}}
  \end{minipage}
 
  \caption{Proton $v_2(p_T)$ (upper panel,
   filled circles) reconstructed from the $"\pi^+"$ candidate sample. 
   The first four points apply to protons directly identified by
   $dE/dx$ (solid asterisks). Statistical errors derive from
   CERES $v_2("\pi^+")$ and $v_2(K_S^0)$ data, and NA49 particle 
   spectra, by Monte-Carlo sampling. Mid-central events, (5.3-14.5)\% 
   of $\sigma_{geo}$, weighted average 9.8\%.
   Lower panel: Systematic errors applicable to data points in upper
   panel, plotted to scale; see Sect.\,\ref{sys}.
  \label{fig:v2prot}}
  \end{figure}
performed by Monte-Carlo simulation of Eq.~\ref{eq:KProtonFlow} using
our $v_2(p_T)$ data for $K^0_S$ and $p_T$ spectra of charged hadrons
from the NA49 Collaboration. The simulation serves to adapt the
NA49-based particle ratios to CERES conditions and not to introduce
any model assumptions.

The resulting proton $v_2(p_T)$ spectrum between 0.30$~$GeV/c and
2.60$~$GeV/c is displayed in Fig.~\ref{fig:v2prot}. It extends over the
range where data both of particle spectra and kaon flow were
available.  At lower $p_T$, four additional points of directly
identified protons\footnote{see $dE/dx$ plot Fig.~\ref{fig:ident_dEdx}
  in Sect.~\ref{pionident}.} are shown by asterisks. The first three of
these data points, from $p_T$\,= 0.05~GeV/c upward, are compatible
with $v_2$= 0, the fourth seems to bridge to the reconstructed points
in an apparently abrupt downward swing. Considering the large
statistical errors, the sawtooth impression may be misleading. The
excursion at low transverse momenta of proton elliptic flow magnitudes
below zero takes its minimum close to 0.4~GeV/c with $v_2$=
-0.0290$\pm~0.0092$, 3.2$\sigma$ below zero.\footnote{weighted mean of
fourth direct and first reconstructed data point in
  Fig.~\ref{fig:v2prot}}.

The systematic errors estimated in Sect.\ref{sys} are displayed in the
bottom panel of Fig.~\ref{fig:v2prot}. The uncertainties due to the
pion flow asymmetry are quantified in the following section
\ref{bound}.

\subsubsection{Setting an upper bound on the flow asymmetry}
\label{bound}

An asymmetry in pion elliptic flow threatens to falsify the results on
proton elliptic flow we derived under the assumption that $v_2$'s are
equal for $\pi^+$ and $\pi^-$. Moreover, the subtraction of
$v_2(\pi^-)$ in place of the smaller $v_2(\pi^+)$ in
Eq.~\ref{eq:KProtonFlow} would be an overcorrection that might have
caused the peculiarity of the proton flow data, i.e. its turn to
negative values at small $p_T$.

\begin{figure}[tt]
  \begin{minipage}{1.0\linewidth} 
  \centerline{\includegraphics[width=6.5cm]
      {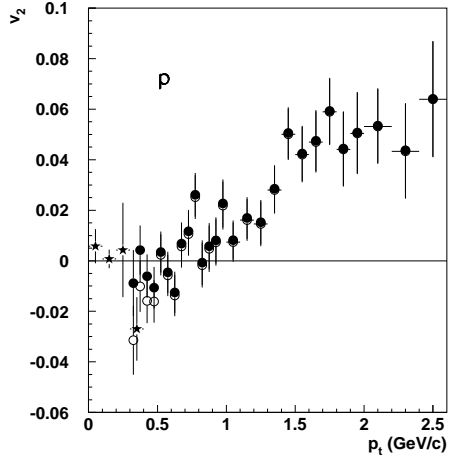}}
  \end{minipage} 
  \caption{Monte-Carlo simulation of proton $v_2$ fixing the asymmetry
  in pion flow ${\cal A}$ at 8.6\% at $p_T$=~0.325 GeV/c such that
  $v_2$(0.325) (the first of the full-circle points) deviates by one
  $\sigma$ from the directly identified point at 0.35~GeV/c (the last
  of the four asterisk points). At higher $p_T$, the ${\cal A}$ values
  are made to decrease rapidly, modelling the strong $p_T$ dependence
  of the preliminary STAR data (full circles). The proton $v_2$ data
  of Fig.~\ref{fig:v2prot} based on ${\cal A}$= 0 are shown by open
  circles. See text.
  \label{fig:asym}}
  \end{figure}

A meaningful {\it upper bound} on the asymmetry, however, is obtained
from the proton $v_2$ data: the lowest of the reconstructed points at
$p_T$= 0.325~GeV/c is close to the $v_2$ point of directly identified
protons at $p_T$=~0.35~GeV/c, marked by an asterisk in
Fig.~\ref{fig:v2prot}; being independent of the asymmetry issue, it
serves as reference point. By tuning the asymmetry parameter in the
Monte-Carlo simulation such as to move the reconstructed point by one
standard deviation {\it above} the reference point,\footnote{speaking
about the {\it rms} of the combined data point errors}, an {\it upper
bound} ${\cal A}_{max}$(0.325)= 8.6\% is obtained. To proceed further,
we turn to the STAR data which show a dramatic decrease of ${\cal A}$
with $p_T$, between 0.3 and 0.5~GeV/c by a factor of $\approx$~3 (see
footnote on p.~24). We use this $p_T$ dependence to extend our upper
bound of asymmetry-related uncertainties beyond the calibration point:
in the simulation shown in Fig.~\ref{fig:asym}, ${\cal A}$ is taken to
decrease linearly to 0.7\% at $p_T$= 0.525~GeV/c and stay constant
above.  We like these calculations be understood as defining the
systematic errors of the proton $v_2(p_T)$ data with respect to the
pion-flow asymmetry issue.

If the asymmetry were caused by $\Delta$ decays, the STAR data scaled
to our centrality would come very close to our bound; and the strong
$p_T$ dependence would be understood. If caused by CME, the scaled 
${\cal A}$ would be much smaller than our upper bound.

\newpage
\subsubsection{The rapidity window for proton flow}
\label{prapi}

\begin{figure}[ht]
  \centerline{\includegraphics[width=6.5cm]{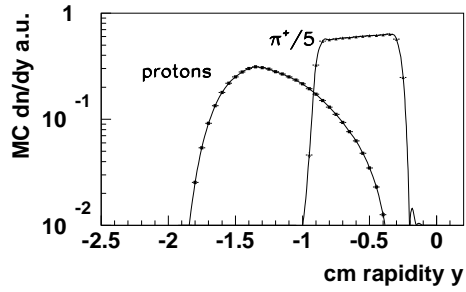}}
  \caption{d$N$/d$y$ of protons and $\pi^+$, in log scale. Monte-Carlo 
   simulation. 
 \label{fig:ypiprot}}
 \end{figure}

Fig.~\ref{fig:ypiprot} demonstrates that protons for which the
elliptic flow data $v_2(p_T$) has been derived originate from about
1.2 units below midrapidity. This reflects the rapidity shift of the
acceptance for low-$p_T$ protons while the $\eta$ acceptance of CERES
can be recognized in the $\pi^+$ distribution (reduced five fold).

\thispagestyle{empty}
\cleardoublepage 

\section{Elliptic flow of $\Lambda$ and $K_{S}^{0}$ particles}
\label{strangeFlow}

\subsection{Results}
\label{KLResults} 

\begin{figure}[ht]
  \centerline{\includegraphics[height=7.5cm] 
{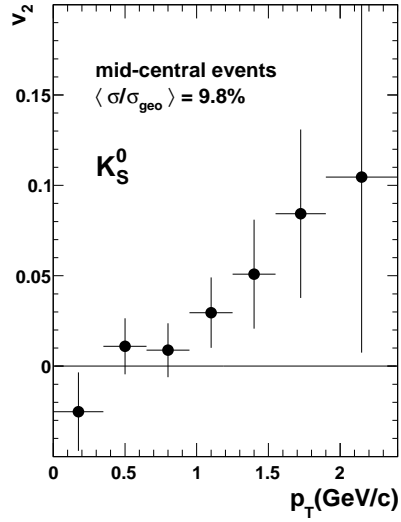}}
  \centerline{\includegraphics[height=7.5cm] 
{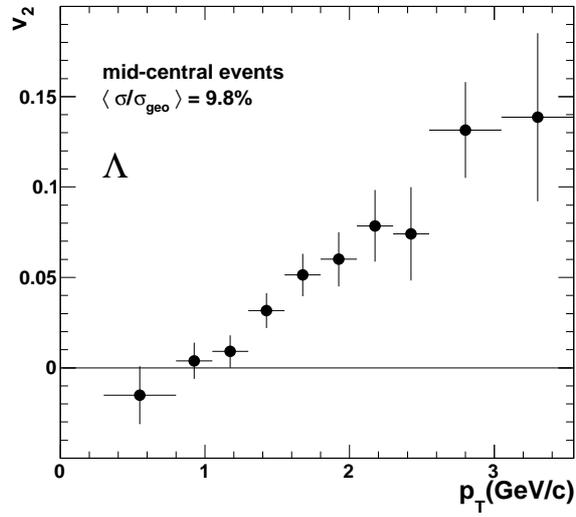}}
  \caption{Differential elliptic flow $v_2(p_T)$ of reconstructed
$K^0_S$ mesons (top) and $\Lambda$ hyperons (bottom). Mid-central
collisions.  Errors are purely statistical.
  \label{fig:exp_Lambda_Kaon}}
  \end{figure}

The upper part of Fig.~\ref{fig:exp_Lambda_Kaon} shows the
differential flow data for $K^0_S$. Statistical errors are large,
mostly due to the strong cut on the secondary vertex position
during reconstruction. We show here the CERES data alone and refer to
a comparison of the combined CERES and NA49 data with hydrodynamical 
calculations to Sect.~\ref{HydroComp}.

Absolute systematic errors in $v_2$ of $K^0_S$, estimated by varying
the cut on the $z$ position of the secondary vertex, are $+0.000 \atop
-0.002$ for $p_{T} < 1.25$~GeV/c and $+0.00 \atop -0.03$ for $p_{T} >
1.25$~GeV/c.  Systematic errors are quadratically small compared to
statistical errors.

The $\Lambda$ elliptic flow displayed in the lower part of
Fig.~\ref{fig:exp_Lambda_Kaon} shows a $p_{T}$ dependence
characteristic for baryon elliptic flow. In the region of small
$p_{T}$, the magnitude of $v_2$ is small but steadily increases with
$p_T$.  At $p_T\approx$ 1.7~GeV/c, $v_2$ exceeds 5\% and rises
further. The absolute systematic error $\Delta v_2$ is estimated from
two different ways of $\Lambda$ reconstruction with emphasis 
either on the size of the signal $S$, or on the signal-to-background
ratio $S/B$; it is $+0.001 \atop -0.007$ for $p_{T} < 1.6$~GeV/c and 
$+0.00 \atop -0.02$ for $p_{T} > 1.6$~GeV/c which is again small 
compared to the statistical errors.

For both species, $K^0_S$ and $\Lambda$, the lowest $v_2$ value at
$p_T$~= 0.175~GeV/c and 0.55~GeV/c, respectively, lies by about
1~$\sigma$ below zero.

\subsection{Comparison to NA49 and STAR experiments}
\label{CompExp} 

\begin{figure}[ht]
  \centerline{\includegraphics[height=8.4cm]
{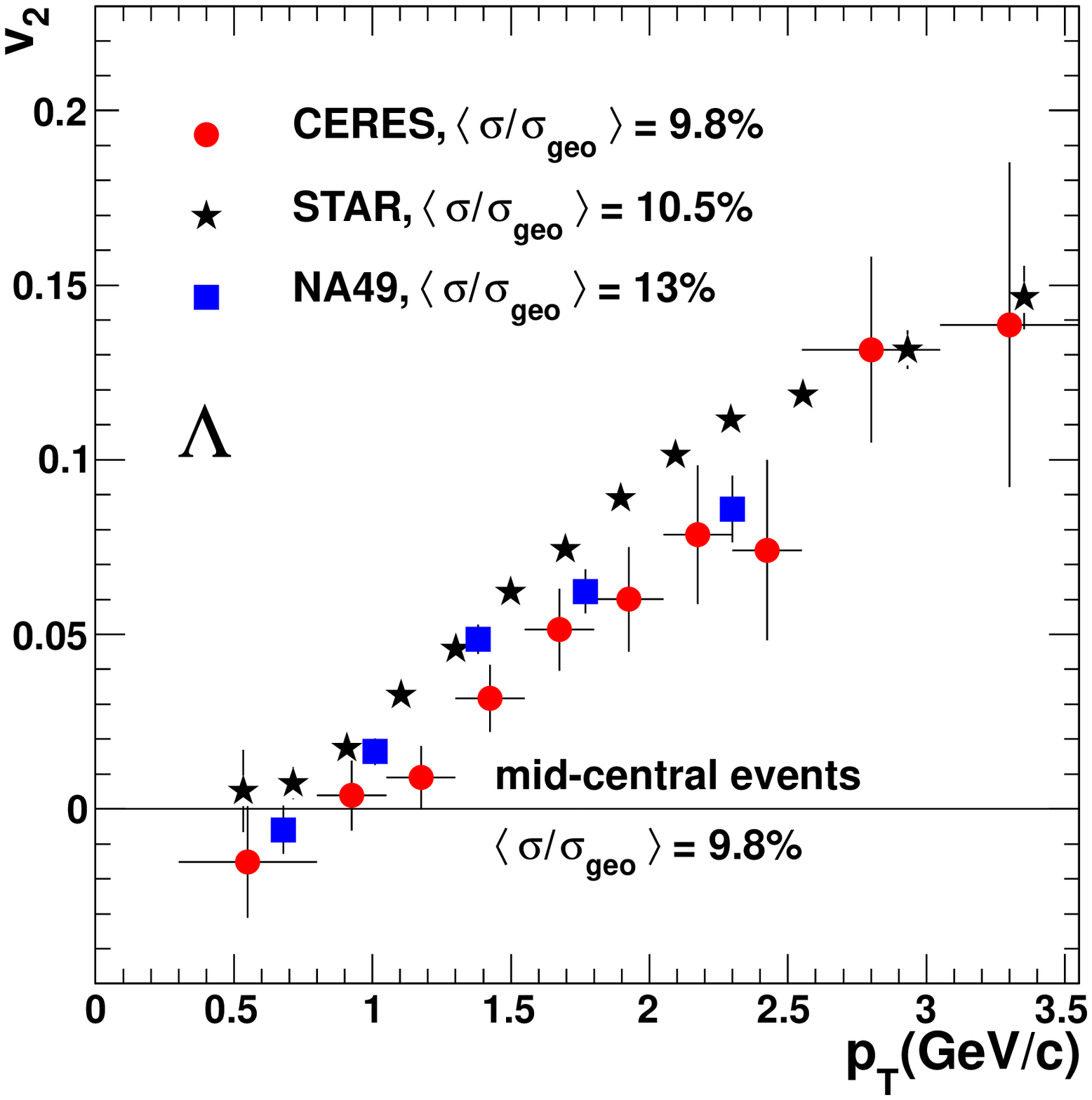}}
  \centerline{\includegraphics[height=8.4cm]
{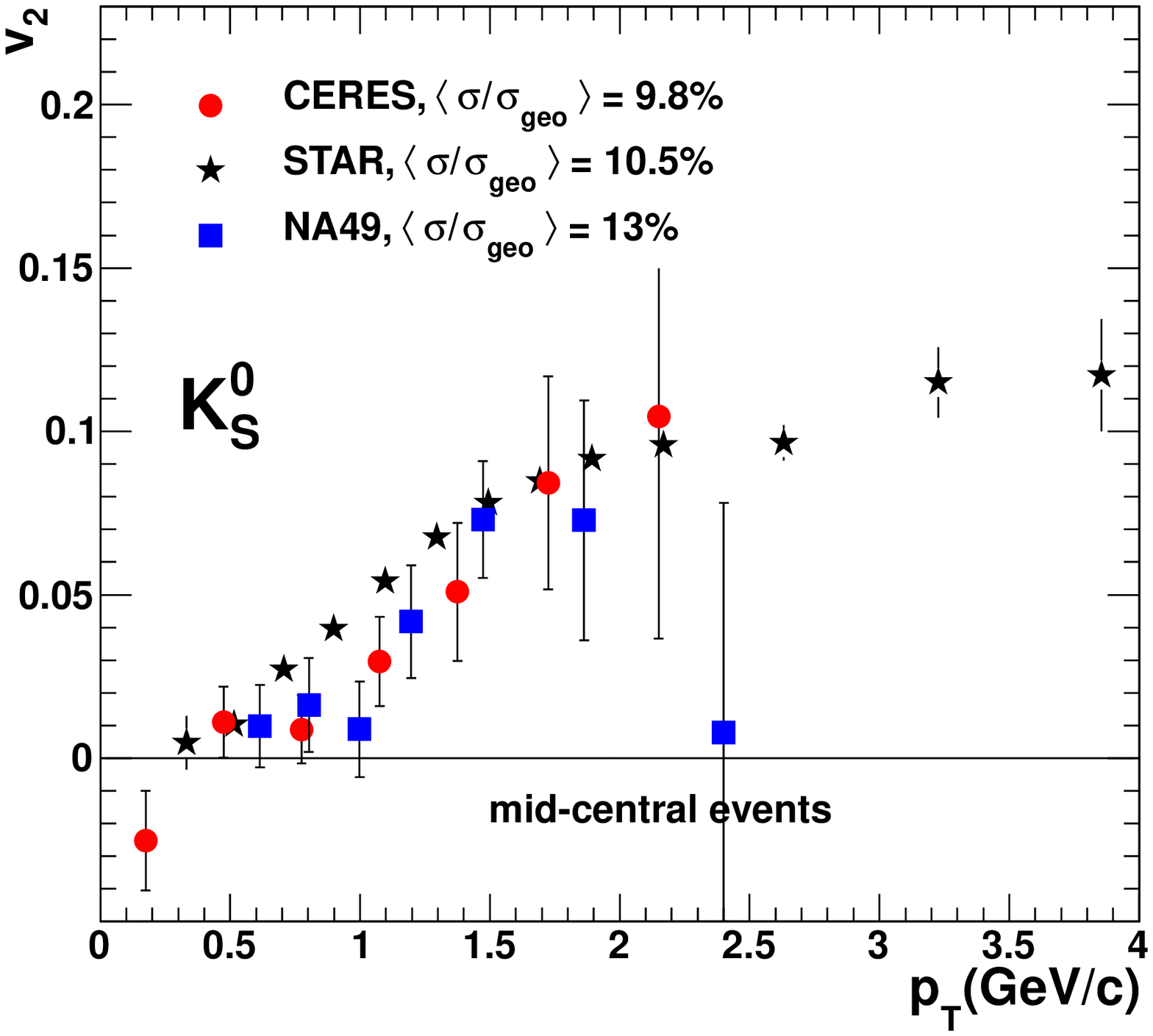}}
  \caption{Comparison of $\Lambda$ (top) and $K^{0}_{S}$ (bottom)
    elliptic flow data of CERES to measurements of NA49, 
    Ref.~\cite{Stefanek06}, and STAR, Ref.~\cite{oldenburg06}.
   \label{fig:compare_RHIC_NA49}}
  \end{figure}

A comparison to results from NA49 \cite{Stefanek06} at the same energy
($\sqrt{s_{NN}}=17.3$~GeV) and to STAR results \cite{oldenburg06} at
$\sqrt{s_{NN}}=200$~Ge is shown in
Fig.~\ref{fig:compare_RHIC_NA49}. The NA49 and CERES data are in
reasonably good agreement.  In order to compare STAR to CERES results,
the former have been rescaled to the centrality used in the CERES
experiment. The appropriate factor is obtained by plotting the STAR
$v_{2}$ values {\it vs} centrality for different transverse momenta of
$\Lambda$ and $K_{S}^{0}$ particles. After rescaling, the STAR the
$v_{2}$ values measured at the RHIC energy are $15-20\%$ higher due to
the higher beam energy.

\thispagestyle{empty}
\cleardoublepage 

\section{Comparison to ideal hydrodynamics calculations}
\label{HydroComp}
\subsection{Overview}
\label{HydroOver}

We compare the results to ideal-hydrodynamics calculations by
P.~Huovinen~\cite{Aga04,Huo05}. These are performed in \mbox{2+1
dimensions} assuming boost-invariant longitudinal flow. Initial
\begin{figure}[tt]
  \centerline{\includegraphics[width=7.50cm,height=4.60cm]
  {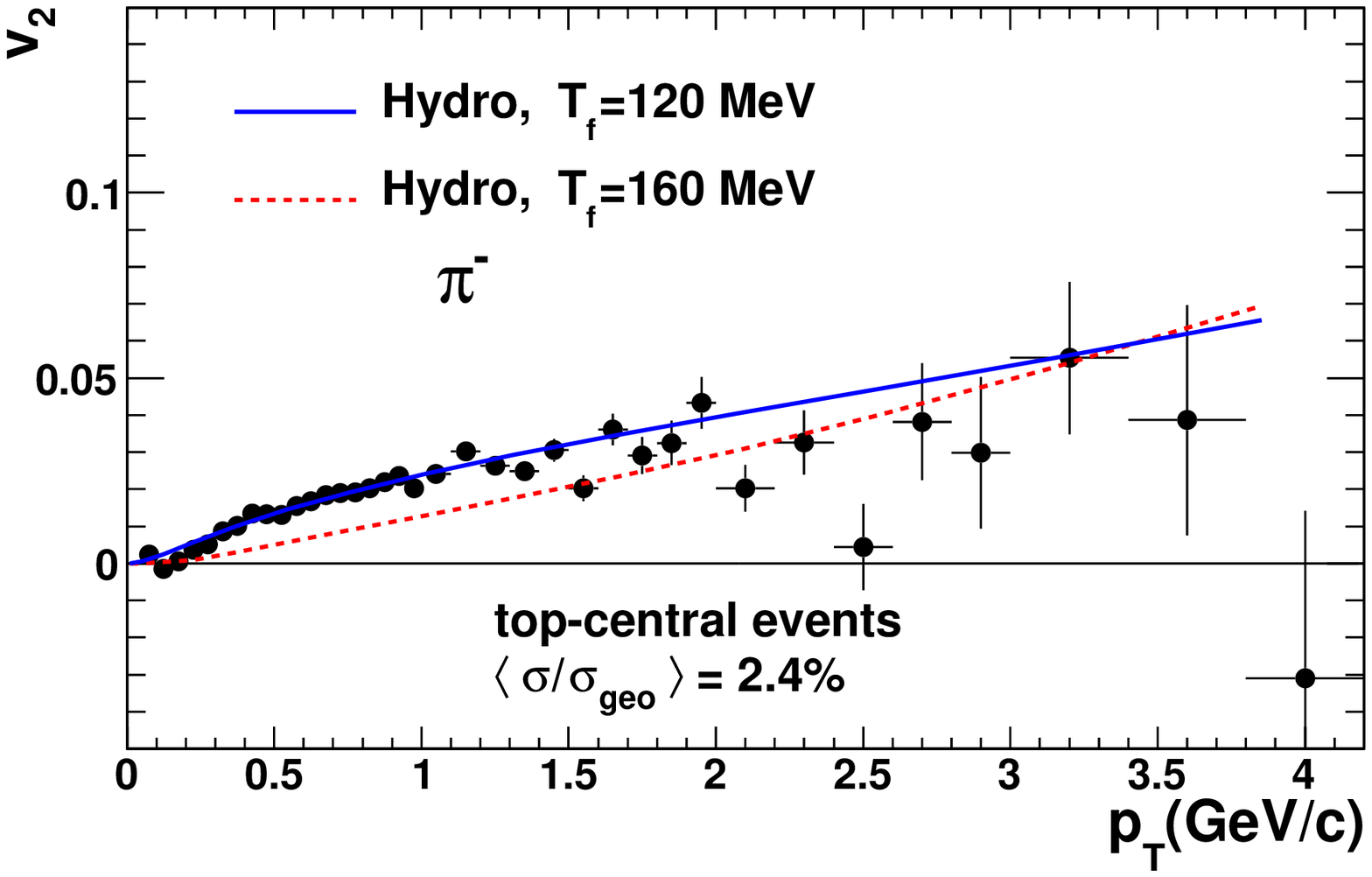}}
  \centerline{\includegraphics[width=7.50cm,height=4.60cm]
  {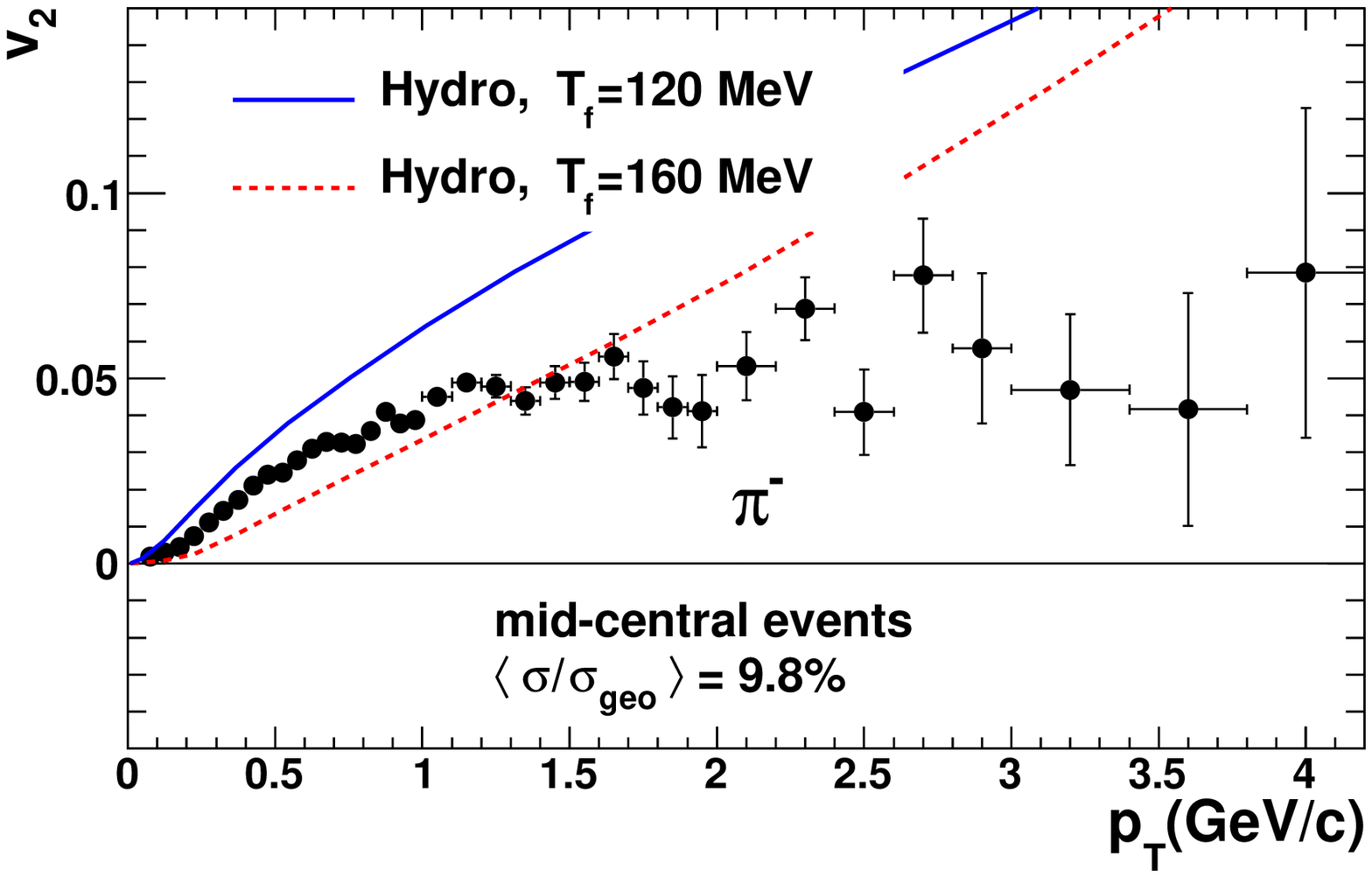}}
 \caption{Pion differential elliptic flow compared to ideal 
  hydrodynamics calculations~\cite{Aga04,Huo05} for two kinetic
  freeze-out temperatures, $T_f$\,= 120~MeV (blue solid) and 
  $T_f$\,= 160~MeV (red dashed). Top: top-central collisions,
  Bottom: mid-central collisions. Data are not corrected for $K^-$
  admixture. Statistical errors.
 \label{fig:pi_hydro}}
 \end{figure}
conditions are fixed by reproducing the $p_T$ spectra of negatively
charged particles and protons in Pb+Pb collisions at top SPS
energy. Assumed is a first order phase transition to quark gluon
plasma at a critical temperature of \mbox{$T_c$\,=~165~MeV}. The
calculations were done for two choices of kinetic freeze-out
temperature\footnote{also known as decoupling temperature $T_{dec}$},
$T_f$\,=~120~MeV and $T_f$\,=~160~MeV.

Kinetic freeze-out at $T_f$\,= 160~MeV may be a handy way to shorten
the evolution of the hadronic fireball and reduce $v_2$ thereby, but
it fails in describing the proton $p_T$ spectra which come out too
steep due to insufficient radial flow, as noticed some time
ago~\cite{Aga04}. 
However, by comparing the data to alternative
freeze-out conditions, we may find out how much the elliptic flow
gains in magnitude during the hadronic fireball evolution, or even
looses.

\subsection{$\pi^-$ elliptic flow}
\label{hydroPions}

The top-central data appear to be in perfect agreement for `standard'
$T_f$\,= 120~MeV, as shown in the upper part of 
Fig.~\ref{fig:pi_hydro}\,\footnote{for easy comparison figures in 
Sect.\,\ref{HydroComp} are plotted to the same scale in $v_2$ and in $p_T$.}.
However, it is hard to rule out, or it may even be likely, that the
good agreement occurs by accidental cancellation between fluctuations
that raise $v_2$ and flow-damping effects reducing it.  For
mid-central collisions, the data seem to prefer a position
in between the two hydro curves till about 1.0~GeV/c and then
saturate, whereas the hydro curves continue to rise about linearly
with $p_T$, see Fig.~\ref{fig:pi_hydro}, bottom.  Note, that
until about 1.2~GeV/c, the mid-central pion flow data remain
significantly {\it above} the $T_f$\,= 160~MeV curve as to be expected.

This is at variance with earlier CERES differential flow
data taken with more peripheral triggers ($\sigma/\sigma_{geo}$=
13\%~-~26\%)~\cite{Aga04}. These elliptic flow data for $h^-$ and
identified $\pi^{\pm}$, the latter with $p_T$ threshold at 1.2~GeV/c,
stay considerably below the predictions for $T_f$\,= 120~MeV, but fall
right on top of the $T_f$\,= 160~MeV line\footnote{referring to Fig.~1c
in \cite{Aga04}.}. In fact, the two CERES data sets seem to confirm
that departures from ideal hydrodynamics increase with increasing
impact parameter of the collision.

Similar conflicts with ideal hydrodynamics have also been reported
from RHIC experiments~\cite{oldenburg06,adams05}. Whether this
indicates incomplete thermalization during primary
stages~\cite{Heinz04,Hir02}, or increasing viscous
corrections~\cite{Teaney03,Shen11}, or a mixture of both, remains to
be seen.
\subsection{$K^0_S$ elliptic flow}
\label{hydroKaons}

\begin{figure}[ttt]
  \centerline{\includegraphics[width=4.30cm,height=5.85cm]
{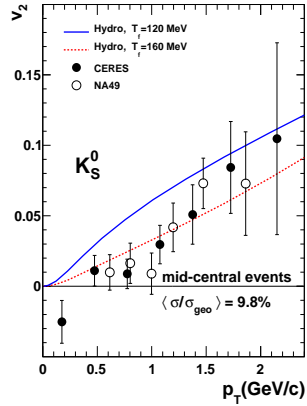}}
  \caption{$K^{0}_{S}$ elliptic flow data from 
  Fig.~\ref{fig:v2-K0S-combined} compared to ideal hydrodynamics
    calculations for $T_{f}$=~120~MeV (blue solid line) and
    $T_{f}$=~160~MeV (red dashed line). 
\label{fig:hydro_K}}
  \end{figure}

The CERES/NA49 combined elliptic flow data for $K^0_S$ are compared in
Fig~\ref{fig:hydro_K} to ideal hydrodynamics calculations.
Surprisingly, for $p_T\leq$\,1.1~GeV/c data points tend to fall
{\it below} the $T_{f}=160$~MeV line.
\begin{figure}[ttt]
  \centerline{\includegraphics[width=4.55cm,height=5.85cm]
{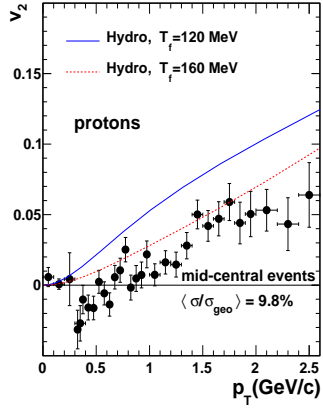}}
  \caption{Proton $v_2(p_T)$ data of Fig.~\ref{fig:v2prot} compared to
ideal hydrodynamics predictions for two freeze-out temperatures (see
Fig.~\ref{fig:pi_hydro}). 
  \label{fig:hydroproton}}
  \end{figure}
\subsection{Proton elliptic flow}
\label{hydroProtons}

The proton elliptic flow $v_2(p_T)$ data are shown again in
Fig.~\ref{fig:hydroproton} for comparison with hydrodynamics
calculations. With standard freeze-out temperature, the calculation
overpredicts the data as in the pion case. What appeared to be a
tendency for $K^0_S$ is for proton $v_2(p_T)$ plain fact: the majority
of data points resides {\it below} the $T_{f}$=~160~MeV line.  With
decreasing $p_T$, the ideal-hydro curve bends smoothly towards zero
while the data continue to fall about linearly until about 0.2~GeV/c.
For such early freeze-out, as
\begin{figure}[bbb]
\centerline{\includegraphics[width=6.3cm,height=5.80cm] 
{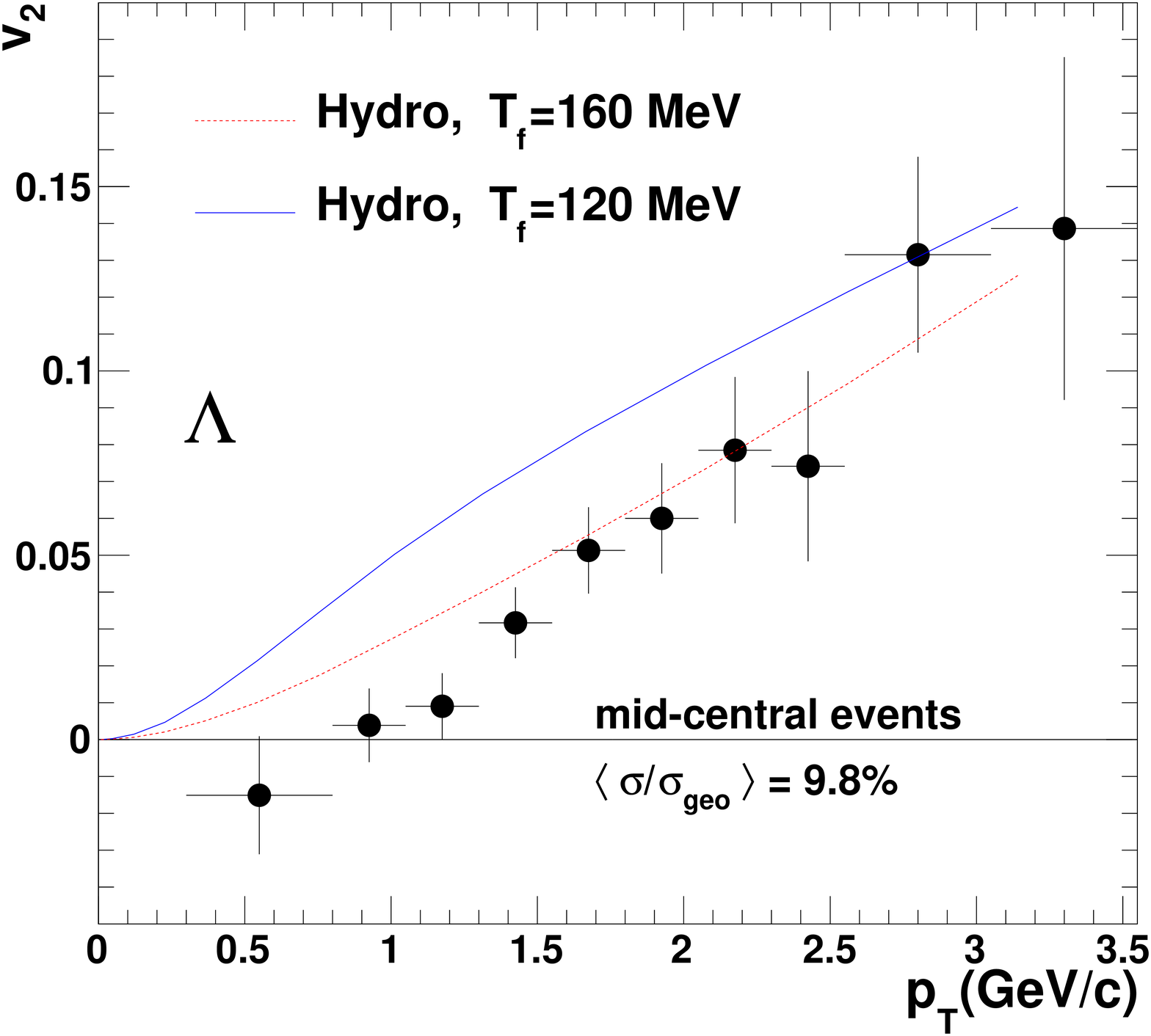}}
  \caption{$\Lambda$ elliptic flow data and predictions of ideal
    hydrodynamics for $T_{f}$\,=~120~MeV (blue solid) and
    $T_{f}$=~160~MeV (red dashed). Mid-central events. Errors
    purely statistical.
\label{fig:hydro_L}}
  \end{figure}
implied by the large $T_{f}$, it is hard to attribute the $v_2$
reduction to a decreased contribution from hadronic interactions. 
The observation rather seems to suggest that a suppression
mechanism or dissipation is at work which is not present in the 
ideal-hydro calculation.

To what extent negative $v_2$ values
are reached is surely a matter of the asymmetry value; for our
upper-bound ${\cal A}$, all negative $v_2$'s with one exception
remain negative. But more important, the perception that $v_2$ data
for protons towards low $p_T$ fall further below and away from the
160~MeV-ideal hydro curve as in the case of $K^0_S$ remains fully
valid. The theory seems to miss some basic ingredient, at least in the
present modelling, to account for the marked reduction in
$v_2$ at low $p_T$ already indicated for $K^0_S$.

\subsection{$\Lambda$ elliptic flow}
\label{hydroLambdas}

The reduction in $v_2$ is seen in Fig.~\ref{fig:hydro_L} to persist
for $\Lambda$ elliptic flow. The $\Lambda$ results could be termed
`perfectly in line' with previous findings, would that not be an
overstatement in view of the large statistical errors and the sparse
low-$p_T$ data in case of the $\Lambda$.  Even an excursion into
negative $v_2$ values as seen for protons would fit into the $\Lambda$
flow data at very low $p_T$.

The data suggest that deviations from ideal hydrodynamics grow with
particle mass. That in fact the deviations seen in $\Lambda$ flow are
not stronger than those in proton flow may well be due to the small
lever arm: the relative gain from $m(p)$ to $m(\Lambda)$ is only 19\%,
compared to the steps $m(\pi)$ to $m(K)$ (250\%), and $m(K)$ to $m(p)$
(90\%); and to limited data precision.

\thispagestyle{empty}
\cleardoublepage 

\section{Discussion and Conclusion}
\label{Discussion}

We have presented differential elliptic flow measurements $v_2(p_T)$
at $\sqrt{s_{NN}}$= 17.3~GeV of $K^{0}_{S}$ and $\Lambda$ in
mid-central collisions, supplemented by $v_2(p_T)$ of negative
pions. We presented also differential elliptic flow $v_2(p_T)$ of
protons, at low $p_T$ by direct identification, for
$p_T\geq$0.35~GeV/c by reconstruction from impure positive-pion
samples. The synopsis of differential elliptic flow spectra from $\pi$
to $K^{0}_{S}$ to $p$ and $\Lambda$ in comparison with calculations of
ideal hydrodynamics discloses a marked decrease of $v_2$ values
towards low $p_T$, getting stronger the larger the hadron masses. That
negative values are reached in proton $v_2$ maybe seen as the most
prominent feature of this trend.

These results were shown to be robust with respect to isospin
asymmetries in elliptic flow of charged pions as reported recently by
STAR at RHIC~\cite{bes11} for minimum-bias $Au+Au$ collisions. An
upper limit on ${\cal A}$ could be derived by using our direct proton
point for reference, at the lowest $p_T$ measured where ${\cal A}$ is
supposedly largest.

Indications of negative-valued elliptic flow of heavy particles from
SPS and RHIC experiments at low $p_T$ and central to mid-central
collisions have been reported~\cite{adams04,alt07,abelev08}, but
differential flow data at SPS or RHIC energies --- with more than one
or two points below zero --- have to our knowledge not been presented
before.  A tendency of massive particles towards negative $v_2$ values
at low $p_T$ is ascribed to the conjunction of elliptic flow and
strong radial expansion under specific freeze-out conditions. The
authors of Ref.~\cite{Huo_plb01} like to illustrate and quantify these
phenomena using a `blast-wave' description~\cite{Schned93} which is
based on an interplay of transverse flow, imparting a momentum gain
proportional to particle mass, and elliptic flow anisotropy which
renders such momentum shift larger in-plane then out-of plane. The
particle density at low $p_T$ is depleted in-plane more than out-of
plane, and elliptic flow values tend to become negative. The depletion
mechanism works best for steep $p_T$ spectra, i.e. those with low
inverse slopes or temperature.

In a recent hydrodynamic study~\cite{Shen11} a large hadronic shear
viscosity-to-entropy ratio was implemented to calculate $p_T$ spectra
and differential elliptic flow of hadrons produced in 200 AGeV Au+Au
collisions.  While viscosity in the late hadronic phase suppresses
elliptic flow in general, the consequences are striking for protons:
for $\eta/s>$~0.42 and $T_f\leq$~120~MeV, proton $v_2$ turns negative
at small $p_T$ and the similarity to our data can hardly be
overseen\footnote{we refer to Fig.~3d and Fig.~4d. of
Ref.~\cite{Shen11}.}. The authors stress that this  effect is entirely
owed to viscous  corrections $\delta f$~\cite{Teaney03} creating large
average  transverse  pressure.  The viscous mechanism, growing  with
$m_T$, appears to initiate a kind  of blast wave with the potential to
produce effects  very similar to  those we observe in  proton elliptic
flow.

Some caution is advised. An earlier but related
viscous hydrodynamics study~\cite{Shen10} remained inconclusive as
proton spectra and charged hadron elliptic flow were found to have
non-compatible requirements on the size of (constant)
$\eta/s$. Besides, Ref.\cite{Molnar11} points to possible consequences
of yet unresolved ambiguities in viscous hydrodynamic calculations for
identified particle observables. 

We hope that our results are useful to better
understand the role of shear viscosity during the late hadronic stages
that terminate heavy-ion collisions at all energies.

\thispagestyle{empty}
\cleardoublepage 

\noindent
{\bf Acknowledgement} \\ 

We are grateful to Christoph Blume and Herbert Str\"obele of the NA49
Collaboration for their advice and for providing us access to
preliminary data. We thank Shusu Shi for communicating the low-energy
STAR data early this year. We enjoyed enlightening discussions with
Pasi Huovinen, appreciate his hydro calculations and are grateful for
his critical reading of the manuscript at an earlier stage.\\






\bibliographystyle{model1-num-names}

\begin{thebibliography}{}
   \bibitem{bra05} I.~Arsene {\it et al}., BRAHMS Collaboration,
      Nucl.~Phys. {\bf A~757}, 1 (2005).      
   \bibitem{pho05} B.B.~Back {\it et al}., PHOBOS Collaboration,
      Nucl.~Phys. {\bf A~757}, 28 (2005).   
   \bibitem{star05} J.~Adams {\it et al}., STAR Collaboration,
      Nucl.~Phys. {\bf A~757}, 102 (2005).  
   \bibitem{phenix05} K.~Adcox {\it et al}., PHENIX Collaboration,
     Nucl.~Phys. {\bf A~757}, 184 (2005).  
   \bibitem{Ollit92} J-Y.~Ollitrault, Phys.~Rev. {\bf D 46}, 229 (1992).
   \bibitem{Bar94} J.~Barrette {\it et al}. E877 Collaboration,
     Phys.~Rev.~Lett. {\bf 73}, 2532 (1994).
   \bibitem{PosVol} A.M.~Poskanzer and S.A.~Voloshin,
     Phys.~Rev. {\bf C~58}, 1671 (1998).
   \bibitem{Apel98} H.~Appelsh\"auser {\it et al}., NA49 Collaboration,
      Phys.~Rev.~Lett. {\bf 80}, 4136 (1998).
   \bibitem{HuoRuu06} P.~Huovinen, P.V.~Ruuskanen, 
      Ann.~Rev.~Nucl.~Part.~Sci. {\bf 56}, 163 (2006).
   \bibitem{GyuMcL05} M.~Gyulassy, L.~McLerran, 
      Nucl.~Phys. {\bf A~750}, 30 (2005).
   \bibitem{HHKLN06} T.~Hirano, U.~Heinz, D.~Kharzeev, R.~Lacey,
      Y.~Nara, Phys.~Lett.{\bf B~636}, 299 (2006).
   \bibitem{Aamodt11} K.~Aamodt {\it et al}., ALICE Collaboration,
      Phys.~Rev.~Lett. {\bf 105}, 252302 (2010).
   \bibitem{Atlas1108} ATLAS Collaboration, Phys.~Lett.{\bf B 707}, 330 (2012).
   \bibitem{Tserruya11} I.~Tserruya, AIP Conf.~Proc. 1422, 166 (2012),
      arXiv:1108.6018.
   \bibitem{Luzum11} M.~Luzum, Phys.~Rev.{\bf C 83}, 044911 (2011).
   \bibitem{SHHS11} C.~Shen, U.~Heinz, P.~Huovinen, H.~Song, 
      Phys.~Rev.{\bf C 84}, 044903 (2011).
   \bibitem{HHN10} T.~Hirano, P.~Huovinen, and Y.~Nara,
      Phys.~Rev.~ {\bf C~83}, 021902 (2011).
   \bibitem{NA4903} C.~Alt {\it et al}., NA49 Collaboration,
      Phys.~Rev.~{\bf C~68}, 034903 (2003). 
   \bibitem{Aga04} G.~Agakichiev {\it et al}., CERES Collaboration,
      Phys.~Rev.~Lett. {\bf 92}, 032301 (2004).  
   \bibitem{Aggar04} M.M.~Aggarwal {\it et al}., WA98 Collaboration, 
      Nucl.~Phys. {\bf A~762}, 129 (2005).
   \bibitem{Huo_plb01} P.~Huovinen, P.F.~Kolb, U.~Heinz, P.V.~Ruuskanen,
      S.A.~Voloshin, Phys.~Lett.{\bf B~503}, 58 (2001).  
   \bibitem{Heinz05} U.~Heinz, J.~.Phys. {\bf G~31}, S717 (2005).
   \bibitem{Heinz04} U.~Heinz, P.~Kolb,J.~Phys. {\bf G~30}, S1229 (2004).
   \bibitem{Teaney03} D.~Teaney, Phys.~Rev. {\bf C~68}, 034913 (2003).
   \bibitem{Niemi11} H.~Niemi, G.S.~Den\-icol, P.~Huovi\-nen, E.~Mol\-nar,
      D.H.~Rischke, Phys.~Rev.~Lett.~{\bf 106}, 212302 (2011).
   \bibitem{Jovan06} J.~Milo\v{s}evi\'c, CERES Collaboration,
       Nucl.~Phys. {\bf A~774}, 503 (2006);\\
       ~~J.~Milo\v{s}evi\'c, Universit\"at Heidelberg, 
       {\it Doctoral Thesis}, 2006.
   \bibitem{Jana03} J.~Slivova (now Biel\v{c}\'{\i}kov{\'a}),
      Universit\"at Heidelberg and Charles University in Prague, {\it
      Doctoral Thesis}, 2003.
   \bibitem{HirGyu_NPA06} T.~Hirano and M.~Gyulassy, 
      Nucl.~Phys. {\bf A~769}, 71 (2006).
   \bibitem{Shen11} C.~Shen, U.~Heinz, 
      Phys.~Rev. {\bf C~83}, 044909 (2011).  
   \bibitem{bes11} S.~Shi for the STAR Collaboration, arXiv:1201.3959;  
      Private Communication S.~Shi, 2011;
      B.~Mohanty for the STAR Collaboration, J.Phys.G:~Nucl.Part.Phys. 
      {\bf 38}, 124023 (2011).
   \bibitem{Marin04} A.~Mar\'{\i}n {\it et al}., CERES Collaboration,
      J.~Phys. {\bf G~30}, S709 (2004).
   \bibitem{NIM08} D.~Adamova {\it et al}., CERES Collaboration, 
      Nucl.~Instr.~Meth. {\bf A~593}, 203 (2008).
   \bibitem{Holl96} P.~Holl, P.~Rehak, F.~Ceretto, U.~Faschingbauer,
      J.P.~Wurm, A.~Castoldi, E.~Gatti,\\
     Nucl.~Instr.~Meth. {\bf A~377}, 367 (1996).
   \bibitem{yurevich06} S.~Yurevich, Universit\"at Heidelberg, {\it Doctoral
       Thesis}, 2006.
   \bibitem{PDG04} S. Eidelman {\it et al}., (Particle Data Group),
     Phys.~Lett. {\bf B~592}, 1 (2004).
   \bibitem{PodAr54} J.~Podolanski and R.~Armenteros, 
     Phil.~Mag. {\bf 45}, 13 (1954).
   \bibitem{Wilrid} W. Ludolphs (now Dubitzky), CERES Collaboration, 
     Heidelberg University, {\it Doctoral Thesis}, 2006.
   \bibitem{Dinh99} P.M.~Dinh, N.~Borghini and J.-Y.~Ollitrault, 
     Phys.~Lett. {\bf B~477}, 51 (2000).
   \bibitem{adamova03} D.~Adamova {\it et al}., CERES Collaboration, 
      Nucl.~Phys. {\bf A~714}, 124 (2003). 
   \bibitem{Tilsner02} H.~Tilsner, CERES Collaboration, 
      Universit\"at Heidelberg, {\it Doctoral Thesis}, 2002.
   \bibitem{BaymBM96} G.~Baym, P.~Braun-Munzinger, Nucl.~Phys. {\bf A~610}, 
       124(1996).
   \bibitem{NA4908} C.~Alt {\it et al}., NA49 Collaboration,
      Phys.~Rev. {\bf C~77}, 034906 (2008); http://na49info.web.cern.ch. 
   \bibitem{Blume08} Ch.~Blume {\it et al}., NA49 Collaboration, 
      J.~Phys.{\bf G~35}, 044004 (2008).
   \bibitem{Stefanek06} G.~Stefanek {\it et al}., NA49 Collaboration, 
       PoS CPD2006:030 [arXiv:nucl-ex/0611003].
   \bibitem{NA4905} C.~Alt {\it et al}., NA49 Collaboration,
      Nucl.~Phys. {\bf A~774}, 473 (2006).
   \bibitem{Blume09} Private communication Ch.~Blume, 2009
   \bibitem{Afanasiev02} S.V.~Afanasiev {\it et al}., NA49 Collaboration,
     Phys.~Rev. {\bf C~66}, 054902 (2002).
   \bibitem{Utvic08} M.~Utvi\'{c}, NA49 Collaboration, Universit\"at Frankfurt,
     {\it Diploma Thesis}, 2006.
   \bibitem{numrec92} Numerical Recipes in C, W.H.~Press, S.A.~Teukolski, 
      W.T.~Vetterling, B.P.~Flannery, Cambridge University Press,
      2nd edition, 1992.
   \bibitem{burnier11} Y.~Burnier, D.E.~Kharzeev, J.~Liao, and H.-U.~Lee,
      Phys.~Rev.~Lett. {\bf 107}, 052303 (2011).
   \bibitem{grecoko05} V.~Greco and C.M.~Ko, Phys.~Rev. {\bf C~71}, 
      041901 (2005).
   \bibitem{molvol03} D.~Molnar and S.A.~Voloshin, 
      Phys.~Rev.~Lett.~{\bf 91}, 092301 (2003).
   \bibitem{nonaka04} C.~Nonaka, B.~M\"uller, M.~Asakawa, S.A.~Bass,
      R.J.~Fries, Phys.~Rev. {\bf C~69}, 031902 (2004). 
   \bibitem{markert07} C.~Markert for the STAR Collaboration, 
      arXiv:0712.1838;\\ 
      C.~Markert, B.~Bellwied, I.~Vitev, Phys.~Lett. {\bf B~669}, 92 (2008).
   \bibitem{dunlop11} J.C.~Dunlop, M.A.~Lisa, and P.~Sorensen (2011),
      Phys.~Rev. {\bf C~85}, 014910 (2012). 
   \bibitem{kharzeev08} D.E.~Kharzeev, L.D.McLerran and H.J.~Warringa,
      Nucl.~Phys. {\bf A~803}, 227 (2008).   
   \bibitem{voloshin04} S.A.~Voloshin, Phys.~Rev. {\bf C~70}, 
      057901 (2004).  
   \bibitem{oldenburg06} M.~Oldenburg {\it et al.}, STAR Collaboration,
     J.~Phys.{\bf G~32}, S563 (2006).
   \bibitem{Huo05} P.~Huovinen (private communications), 2005.
   \bibitem{adams05} J.~Adams {\it et al}., STAR Collaboration,
      Phys.~Rev. {\bf C~72}, 014904 (2005).
   \bibitem{Hir02} T.~Hirano, Phys.~Rev. {\bf C 65}, 011901(R) (2002).   
   \bibitem{adams04} J.~Adams {\it et al}., STAR Collaboration,
      Phys.~Rev.~Lett. {\bf 92}, 052302 (2004).
  \bibitem{alt07} C.~Alt {\it et al}., NA49 Collaboration,
      Phys.~Rev. {\bf C~75}, 044901 (2007). 
   \bibitem{abelev08} B.I.~Abelev {\it et al}., STAR Collaboration,
      Phys.~Rev. {\bf C~77}, 054901 (2008).
   \bibitem{Schned93} E.~Schnedermann, J.~Sollfrank, U.~Heinz,
      Phys.~Rev. {\bf C~48}, 2462 (1993).   
   \bibitem{Shen10} C.~Shen, U.~Heinz, P.~Huovinen, H.~Song, 
      Phys.~Rev. {\bf C~82}, 054904 (2010).
   \bibitem{Molnar11} D.~Molnar, J.~Phys. {\bf G~38}, 124173 (2011).   

\end{thebibliography}



\end{document}